\documentclass[preprint,times,10pt,twocolumn]{elsarticle}

\usepackage[numbers]{natbib}
\usepackage{amsmath,amssymb,amsfonts}
\usepackage{algorithmic}
\usepackage{graphicx}
\usepackage{textcomp}
\usepackage{xcolor}
\usepackage{amsmath}
\usepackage{amsfonts}
\usepackage{tikz}
\usetikzlibrary{shapes.geometric, arrows}
\usepackage{subcaption}
\usepackage{algorithm}
\usepackage{array} 
\usepackage{booktabs} 
\usepackage{multirow}

\usepackage{cleveref}
\Crefname{figure}{Fig}{Figures}

\journal{Pattern Recognition}

\begin{document}

\begin{frontmatter}



\title{Backdoor Attacks against Hybrid Classical-Quantum Neural Networks}

\author[1]{Ji Guo}
\ead{jiguo0524@gmail.com}

\affiliation[1]{organization={Laboratory Of Intelligent Collaborative Computing, University of Electronic Science and Technology of China},
    city={Chengdu},
    postcode={611731}, 
    country={China}}


\author[2]{Wenbo Jiang\corref{cor1}}
\ead{wenbo_jiang@uestc.edu.cn}

\affiliation[2]{organization={School of Computer Science and Engineering, University of Electronic Science and Technology of China},
    city={Chengdu},
    postcode={611731}, 
    country={China}}

\affiliation[3]{organization={School of Computer Science and Artificial Intelligence, Wuhan University of Technology},
    city={Wuhan},
    postcode={430070}, 
    country={China}}
    
\author{Rui Zhang \fnref{2} }
\author{Wenshu Fan \fnref{2} }
\author{Jiachen Li \fnref{3} }
\author{Guoming~Lu \fnref{1} }

\cortext[cor1]{Corresponding author}

\begin{abstract}

Hybrid Quantum Neural Networks (HQNNs) represent a promising advancement in Quantum Machine Learning (QML), yet their security has been rarely explored. In this paper, we present the first systematic study of backdoor attacks on HQNNs. We begin by proposing an attack framework and providing a theoretical analysis of the generalization bounds and minimum perturbation requirements for backdoor attacks on HQNNs. Next, we employ two classic backdoor attack methods on HQNNs and Convolutional Neural Networks (CNNs) to further investigate the robustness of HQNNs. Our experimental results demonstrate that HQNNs are more robust than CNNs, requiring more significant image modifications for successful attacks. Additionally, we introduce the Qcolor backdoor, which utilizes color shifts as triggers and employs the Non-dominated Sorting Genetic Algorithm II (NSGA-II) to optimize hyperparameters. Through extensive experiments, we demonstrate the effectiveness, stealthiness, and robustness of the Qcolor backdoor.
\end{abstract}


\end{frontmatter}

\maketitle

\section{Introduction}
\label{Introduction}

Hybrid Classical-Quantum Neural Networks (HQNNs) \cite{Farhi2018Classification, Zhao2021QDNN:, Andrea2020Transfer}, a widely used Quantum Machine Learning (QML) \cite{Biamonte2017Quantum} method, have achieved success in many tasks such as image classification \cite{Sebastianelli2022On}, generative modeling \cite{Zoufal2019Quantum}, and reinforcement learning \cite{Jerbi2021Parametrized}. These models combine classical neural networks with Quantum Neural Networks (QNNs) \cite{Beer2020Training} and utilize optimized classical algorithms \cite{SGD,ADAM} for training. Compared to classical deep neural networks (DNNs) \cite{LeCun2015Deep}, HQNNs take advantage of quantum computing to provide faster processing speeds and more complex probability distributions. Moreover, HQNNs are more practical to implement than QNNs, as they require fewer quantum bits, making them a more feasible solution given the current limitations in quantum hardware technology.

While QML models have achieved excellent performance in various domains, recent attention has shifted towards their safety \cite{Liao2021Robust, West2023Towards, Ren2022Experimental, Chu2023QDoor:, Chu2023QTrojan:}. Drawing inspiration from security threats to DNNs, most studies have focused on adversarial attacks \cite{Szegedy2013Intriguing} on QML \cite{Liao2021Robust, West2023Towards, Ren2022Experimental}. However, backdoor attacks, another well-known threat to DNNs \cite{Gu2019BadNets:}, have seldom been explored for QML. Backdoor attacks \cite{Gu2019BadNets:, Li2020Backdoor} pose a significant threat by inserting malicious samples with specific triggers into the training dataset. After training, the model performs accurately on clean inputs but misclassifies triggered inputs as pre-defined labels. To our knowledge, there is no work elaborated on backdoor attacks on HQNNs. 
Given the distinct classification principles of HQNNs, which involve using both classical and quantum state data and utilizing Hilbert space for classification, a critical question arises regarding the robustness of HQNNs against backdoor attacks.

\begin{figure*}[ht]
\centerline{\includegraphics[width=0.8\linewidth]{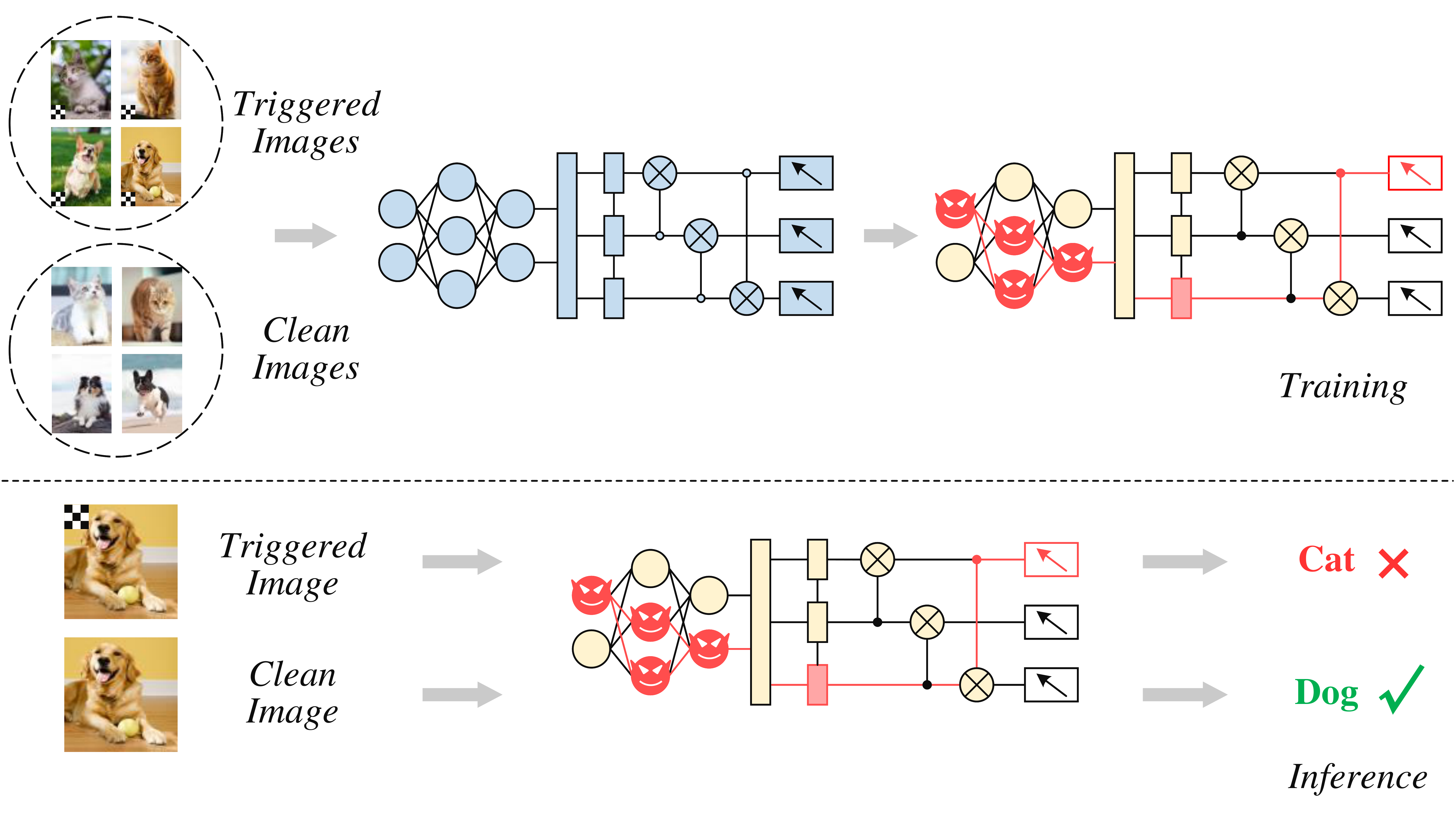}}
\caption{Overview of backdoor attacks on HQNN}
\label{overview of hqnn backdoor}
\end{figure*}

To fill this gap, We evaluate the robustness of HQNNs under two basic backdoor triggers with various trigger settings. Fig.~\ref{overview of hqnn backdoor} illustrates an overview of backdoor attacks on HQNN. Our experimental results demonstrate that HQNNs exhibit better robustness against backdoor attacks than Convolutional Neural Networks (CNNs) \cite{resnet,LeCun2015Deep}, and the effectiveness of backdoors in HQNNs also relies on the features extracted by CNN lays.
Moreover, the robustness of HQNNs against backdoor attacks presents two challenges when adapting CNN-based backdoor attacks to HQNNs: (1) the attacks require more significant modifications, which reduces stealthiness; and (2) the attacks require more poisoned samples to learn the trigger patterns, making success difficult at low poisoning rates.

\begin{figure}[ht]
    \centering
    \begin{subfigure}{1\linewidth}
        \includegraphics[width=1\linewidth]{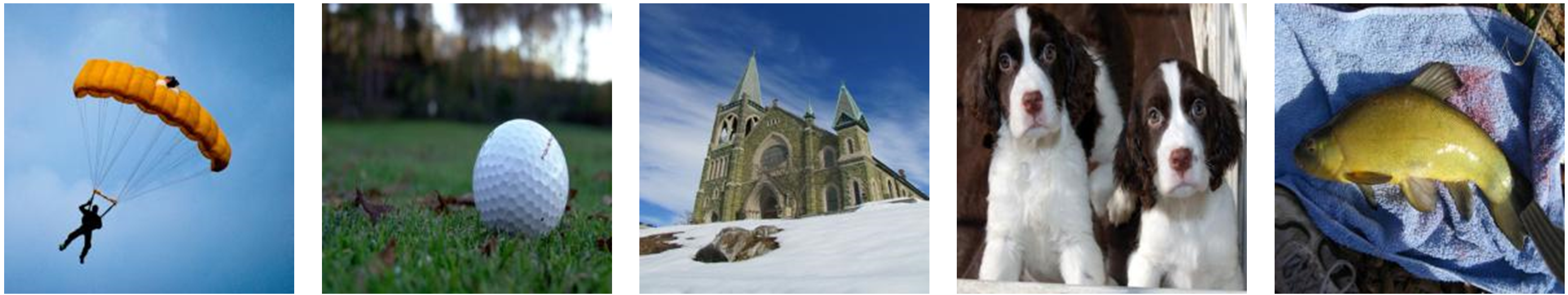}
        \caption{Clean images}
    \end{subfigure}
    \hfill
    \begin{subfigure}{1\linewidth}
        \includegraphics[width=1\linewidth]{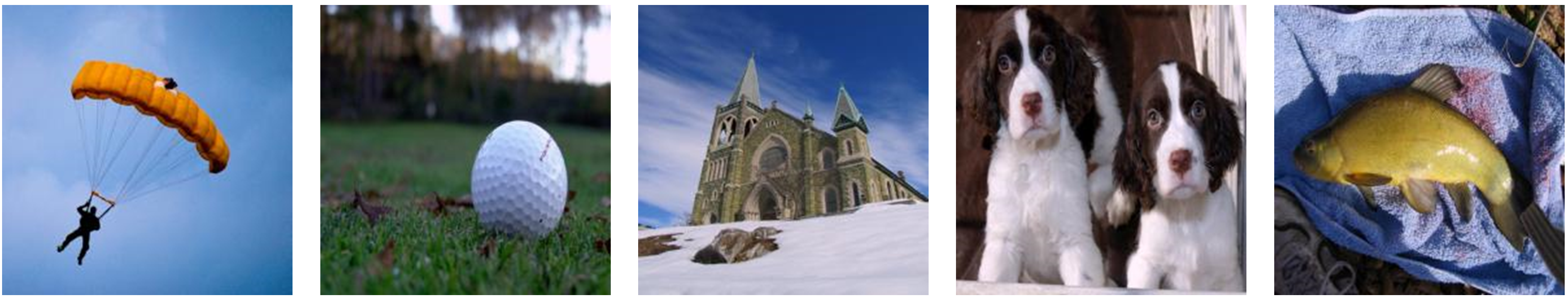}
        \caption{Qcolor backdoor triggered images}
    \end{subfigure}

    \caption{Example of Qcolor backdoor triggered images and clean images from Imagenette}
    \label{plt}
    
\end{figure}

To achieve a stealthy and low poisoning rate backdoor attack in HQNNs, we introduce a novel backdoor attack named Qcolor backdoor. Inspired by adversarial attacks on HQNNs, we consider that Variational Quantum Circuits (VQCs) \cite{Biamonte2017Quantum} encode qubit based on image colors for designing backdoor attacks. We adjust the ratios of the three color channels, using color shifts as the trigger, and employ the Non-dominated Sorting Genetic Algorithm II (NSGA-II) \cite{Deb2002A} algorithm to find the optimal hyperparameters to ensure both effectiveness and stealthiness. A subtle overall color shift can change the VQC’s encoding angle, classifying triggered images into a specified class. Color is a significant feature for CNNs and can be easily learned. Moreover, the human visual system focuses more on the gradient of color changes rather than the absolute value of the overall color, making subtle overall color changes likely to go unnoticed.

Although existing methods use the superimposition of a fixed color layer as a trigger to attack CNNs \cite{jiang2023color}, this approach requires more significant color changes to successfully attack HQNNs. This reduces the stealthiness of color backdoor in HQNNs, making anomaly detection easier. Additionally, our experiments show that the method of superimposing a color layer in CNNs fails to successfully attack HQNNs at low poisoning rates.
Compared to traditional color backdoors, our method achieves a high Attack Success Rate (ASR) while maintaining a high Structural Similarity Index Metric (SSIM) \cite{Wang2004Image}, and it can successfully attack with low poisoning rates. Fig.~\ref{plt} shows the clean images and Qcolor backdoor triggered images. 

Our main contributions are as follows:
\begin{itemize}
    \item 
    We provide a framework and theoretical analysis of backdoor attacks on HQNNs and systematically explore the robustness of HQNNs against backdoor attacks.
    \item 
    We propose the Qcolor backdoor, a novel backdoor attack for HQNNs that adaptively adjusts the rates of the three color channels and uses color shifts as the trigger. To enhance stealthiness while maintaining the effectiveness of the attack, we employ the NSGA-II algorithm to optimize the hyperparameters of the Qcolor backdoor. 
    \item 
    We evaluate robustness of Qcolor backdoor against three SOAT backdoor defenses from CNNs: STRIP, Neural Cleanse, and Fine-Pruning.
\end{itemize}

The remainder of this paper is organized as follows. In Section \ref{Related Work}, we review the existing studies related to this work. In Section \ref{Threat Model}, we introduce our threat model. In Section \ref{Backdoor Attacks in HQNNs},  we present the general framework and theoretical analysis of backdoor attacks on HQNNs. In Section \ref{The Proposed Backdoor Attacks: Qcolor Backdoor}, we introduce the Qcolor backdoor. In Section \ref{Experiments}, we experiment on the robustness of HQNNs against backdoor attacks and the performance of the Qcolor backdoor. In Section \ref{Conclusion}, we summarize our work.

\section{Related Work}
\label{Related Work}

\subsection{Hybrid Classical-Quantum Neural Networks}
Hybrid Classical-Quantum Neural Networks (HQNNs) \cite{Farhi2018Classification, Andrea2020Transfer} integrate classical neural networks with quantum neural networks, harnessing quantum advantages while minimizing the need for quantum qubits. This integration has demonstrated success in various tasks, including image classification \cite{Sebastianelli2022On}, generative modeling \cite{Zoufal2019Quantum}, and reinforcement learning \cite{Jerbi2021Parametrized}.

In this work, we focus on image classification, a fundamental task in computer vision. The classical neural network component, typically convolutional neural networks (CNNs) \cite{LeCun2015Deep, resnet}, extracts feature maps from images. These feature maps are then processed by the quantum neural network component, which employs variational quantum circuits (VQC) \cite{Biamonte2017Quantum}. The VQC is composed of a quantum encoder, a variational model, and measurement components (see Fig.~\ref{Overview of Qcolor backdoor} (c)).

This hybrid approach combines the powerful feature extraction capabilities of CNNs with the computational potential of quantum circuits, aiming to enhance the performance and efficiency of image classification models.

\subsection{Security in Quantum Machine Learning}

In recent years, a growing body of research has focused on the security aspects of quantum machine learning (QML), with significant emphasis on adversarial attacks \cite{West2023Towards,Ren2022Experimental,Du2021Quantum,Gong2024Enhancing,Liu2019Vulnerability}. Early studies by Liu et al. \cite{Liu2019Vulnerability} highlighted that QML methods are particularly vulnerable to adversarial attacks due to a geometric property known as the concentration of measure phenomenon (COMP) in Hilbert spaces. This property, relevant to spaces used for classification tasks in quantum computing, makes QML models susceptible regardless of the specific classifier details.

Further research has explored various aspects of QML security. For instance, Wendlinger et al. discussed using Lipschitz bounds to evaluate the robustness of quantum models, demonstrating that tight bounds can enhance the robustness of quantum circuits \cite{Wendlinger2024A}. Additionally, studies have examined different data encoding techniques, such as data re-uploading, angle encoding, and amplitude encoding, and their impacts on model robustness \cite{Gong2024Enhancing}.

Despite the focus on adversarial attacks, relatively less attention has been paid to backdoor attacks in QML. To date, the primary research on backdoor attacks in QML has been conducted by Chu et al., who explored these attacks in the context of VQC but did not address hybrid quantum-classical HQNNs \cite{Chu2023QDoor:, Chu2023QTrojan:}. Due to the structural differences between HQNNs and QNNs, their work cannot be directly applied to HQNNs. Therefore, investigating the robustness of HQNNs against backdoor attacks is a valuable research question.

\subsection{Backdoor Attacks and Defences}

Backdoor attacks \cite{Gu2019BadNets:,Li2020Backdoor} pose a significant threat to DNNs. These attacks involve embedding malicious samples with specific triggers into the training dataset. After the model is trained with these triggered data, it performs correctly on clean inputs but misclassifies triggered inputs. The pioneering study, BadNets \cite{Gu2019BadNets:}, used a white patch as a trigger to attack CNNs. Subsequently, Liu et al. \cite{Saha2020Hidden} introduced clean-label backdoor attacks. Building on these foundational works, researchers have further explored invisible backdoor attacks by applying different techniques to triggers \cite{Doan2021Backdoor,jiang2023color,Nguyen2021WaNet}, using clean labels for backdoor inputs \cite{Saha2020Hidden}, and manipulating the training process \cite{Eugene2021Blind}. Additionally, some studies have achieved backdoor attacks without data poisoning by directly modifying model parameters \cite{Dumford2018Backdooring} and using Trojan implants \cite{Tang2020An}.

To deal with those attacks, various backdoor defense methods have been developed \cite{Liu2022Backdoor,Zhang2023Backdoor,Zhu2023Neural,Gao2019STRIP:}. One of the most influential methods is Neural Cleanse \cite{Wang2019Neural}, which is widely used for backdoor detection and identification. This technique generates triggers for backdoored models through reverse engineering in black-box scenarios and identifies the attacker's target class by comparing the anomaly indices of different classes. Another approach, the pruning-based defense method \cite{liu2018finepruning}, targets and suppresses neurons responsible for backdoor behavior within the compromised model, effectively neutralizing the threat by selectively disabling the affected neurons. Additionally, Stronghold Testing of Regular Input Pathways (STRIP) \cite{Gao2019STRIP:} detects potential backdoor by repeatedly perturbing the same input image and observing the consistency of the model's outputs.

\section{Threat Model}
\label{Threat Model}

We adopt the threat model with numerous backdoor attacks on CNNs prior studies \cite{Gu2019BadNets:,jiang2023color,Nguyen2021WaNet,Liu2020Reflection}. 
Our approach involves generating triggered samples without the target class and incorporating them into the clean training dataset before releasing them publicly. A victim developer inadvertently introduces a backdoor vulnerability upon using this tampered dataset to train their model. It is important to note that the attacker is presumed to have neither control over the training process nor any knowledge about the specifics of the victim's model. Typically, our backdoor attacks have the following goals:
\begin{itemize}
    \item \textit{Functionality-preserving}. Test accuracy of clean samples in the backdoor model should have a minor impact. 
    
    \item \textit{Effectiveness.} Most of the triggered samples should be classified into the target class. 
    \item \textit{Stealthiness.} The triggered sample should be similar to clean samples and could be natural-looking to human eyes.
\end{itemize}

\section{Backdoor Attacks in HQNNs}

\label{Backdoor Attacks in HQNNs}
\subsection{Formulation of Backdoor Attacks in HQNNs}
\label{Formulation of Backdoor Attacks in HQNNs}
Backdoor attacks embed malicious triggers within a model to perform normally on clean data but exhibit targeted behavior when a specific trigger is present. Compared to backdoor attacks in CNN, HQNN backdoor attacks involve injecting triggers into both the CNN and QNN components.

Let \(\mathcal{D} = \{(x_i, y_i)\}_{i=1}^n\) denote the clean training dataset, where \(x_i\) is an input sample and \(y_i\) is the corresponding label. Let \(\mathcal{D}_t = \{(x_i', y_i')\}_{i=1}^m\) be the triggered dataset with a trigger embedded in the input samples. Let \(f_{HQ}(x) = f_Q(f_C(x))\) represent the HQNN model, where \(f_C\) and \(f_Q\) denote the classical and quantum components, respectively. 

The goal of a backdoor attack against HQNN can be formulated:
\begin{itemize}
    \item For clean data \(x \in \mathcal{D}\), the model behaves normally: \(f_{HQ}(x) \approx y\).
    \item For triggered data \(x' \in \mathcal{D}_t\), the model outputs the attacker-specified label: \(f_{HQ}(x') = y_t\), where \(y_t\) is the target label.
\end{itemize}

The clean training objective function can be described as follows:
\begin{equation}
\min_{\theta_C, \theta_Q} \sum_{(x, y) \in \mathcal{D}} \mathcal{L}_c(f_{HQ}(x; \theta_C, \theta_Q), y)
\end{equation}
where \(\mathcal{L}_c\) is the loss function for clean data, and \(\theta_C\) and \(\theta_Q\) are the parameters of the CNN and QNN components, respectively.

The triggered training objective function can be described as follows:
\begin{equation}
\min_{\theta_C, \theta_Q} \sum_{(x', y_t) \in \mathcal{D}_t} \mathcal{L}_t(f_{HQ}(x'; \theta_C, \theta_Q), y_t)
\end{equation}
This ensures that the model misclassifies triggered data as the target label \(y_t\). So the total objective function is:
\begin{equation}
\begin{split}
&\min_{\theta_C, \theta_Q} \left( \sum_{(x, y) \in \mathcal{D}} \mathcal{L}_c(f_{HQ}(x; \theta_C, \theta_Q), y) + \lambda \sum_{(x', y_t) \in \mathcal{D}_t} \mathcal{L}_t(f_{HQ}(x'; \theta_C, \theta_Q), y_t) \right)
\end{split}
\end{equation}

Considering the difference in gradient calculations for VQC in QNNs compared to traditional neural networks, the total gradient update during training can be described as follows:
\begin{equation}
\theta \leftarrow \theta - \eta \left( \nabla_{\theta} \mathcal{L}_c + \lambda \nabla_{\theta} \mathcal{L}_t \right)
\end{equation}
where \(\theta = (\theta_C, \theta_Q)\) represents the parameters of both the CNN and QNN components and \(\eta\) is the learning rate.

\subsection{Theoretical Analysis for Backdoor Attacks in HQNN}

\label{Theoretical Analizy for Backdoor Attacks in HQNN}

In this section, we provide a theoretical analysis of two aspects: the generalization lower bound and the minimum trigger perturbation required for feature distribution changes. The proof is provided in the Appendix.

\subsubsection{Generalization Lower Bound}

Let $\mathcal{H}$ be a Hilbert space, $\mathcal{D}_t = \{(x_i', y_i')\}_{i=1}^m$ be the dataset with embedded triggers, $f_{HQ}$ be the HQNN model, and $\mathcal{L}_t$ be the loss function for the triggered data. 
We define training error on triggered samples as:
\begin{equation}
\hat{R}_t(f_{HQ}) = \frac{1}{m} \sum_{i=1}^m \mathcal{L}_t(f_{HQ}(x_i'), y_i')
\end{equation}
The generalization error on triggered samples as:
\begin{equation}
R_t(f_{HQ}) = \mathbb{E}_{(x', y') \sim P_t}[\mathcal{L}_t(f_{HQ}(x'), y')]
\end{equation}

To obtain the generalization lower bound for HQNN under backdoor attacks, we give the following regularization conditions:

\begin{enumerate}

\item The norm of the output of the HQNN model $f_{HQ}$ is bounded by a constant $B$ for all inputs in the triggered dataset $\mathcal{D}_t$:
\begin{equation}
    \|f_{HQ}(x)\| \leq B, \quad \forall x \in \mathcal{D}_t
\end{equation}
    
\item The loss function $\mathcal{L}_t$ is Lipschitz continuous with Lipschitz constant $L_t$:
\begin{equation}
\begin{split}
    |\mathcal{L}_t(f_{HQ}(x'), y) - \mathcal{L}_t(f_{HQ}(x), y)| \leq L_t \|x' - x\|
\end{split}
\end{equation}
    
\item The triggered sample $x'$ can be expressed as a linear combination of $x$ and the trigger pattern $z$, with $\delta$ denoting the trigger strength:
\begin{equation}
    x' = x + \delta z
\end{equation}

\end{enumerate}

\textbf{Theorem 1.}
If conditions 1 to 3 are satisfied, then the generalization lower bound for HQNN under backdoor attacks satisfies:
\begin{equation}
\begin{split}
\label{Theorem 1}
R_t(f_{HQ}) \geq \hat{R}_t(f_{HQ}) - \frac{B}{\sqrt{2m}} \sqrt{\ln \frac{2}{\delta}} + L_t \delta \|z\|
\end{split}
\end{equation}

According to Equation (\ref{Theorem 1}), the number of triggered samples \(m\) plays a significant role; as \(m\) increases, the term \(\frac{B}{\sqrt{2m}} \sqrt{\ln \frac{2}{\delta}}\) decreases, reducing the gap between the generalization error and the training error. Consequently, a larger triggered sample size leads to a higher ASR.
In addition, the trigger strength \(\delta\) is directly proportional to the term \(L_t \delta \|z\|\), meaning that stronger triggers result in a higher generalization error, thereby increasing the ASR. 
Besides, the Lipschitz continuity constant \(L_t\) of the triggered loss function also significantly influences the generalization error. A larger \(L_t\) value implies a higher term \(L_t \delta \|z\|\), indicating that the smoothness of the loss function affects the model's robustness to backdoor attacks.

\subsubsection{Minimum Trigger Perturbation for Feature Distribution Change }

To prove the robustness of HQNNs under backdoor attacks, we need to determine the minimum perturbation strength \(\delta\) required to change the feature distribution significantly. We will analyze this using the concentration of measure phenomenon.\cite{Ledoux2005The}.

We define the perturbed feature distribution as:
\begin{equation}
\phi_\delta(x) = \phi(x + \delta)
\end{equation}
where we want to show that the perturbed feature \(\phi_\delta(x)\) remains concentrated around \(\mathbb{E}[\phi(x)]\).

\textbf{Theorem 2.}
If the same conditions 1 to 3 are satisfied, then the minimum perturbation strength \(\delta\) required for backdoor attacks in HQNNs satisfies:
\begin{equation}
\|\delta\| \geq c^{-1}(\epsilon)
\label{MinTriggerPerturbation}
\end{equation}
where \(c^{-1}(\epsilon)\) is the inverse function of \(c(\epsilon)\), \(\epsilon\) represents the allowed deviation, usually a small positive number, and \(c(\epsilon)\) is a function such that:
\begin{equation}
\mu(\{x \in S : \|\phi(x) - \mathbb{E}[\phi(x)]\| \geq \epsilon\}) \leq e^{-c(\epsilon)}
\end{equation}

Based on the above conditions and Equation (\ref{MinTriggerPerturbation}), to significantly change the feature distribution of HQNNs and successfully perform a backdoor attack, the required trigger strength \(\delta\) must be at least \(c^{-1}(\epsilon)\), which means the strength \(\delta\) required for a successful trigger injection grows exponentially with \(c\). This demonstrates that VQC can enhance the model's robustness against backdoor attacks.

\section{The Proposed Backdoor Attack: Qcolor Backdoor}
\label{The Proposed Backdoor Attacks: Qcolor Backdoor}

\begin{figure*}[ht]

\centerline{\includegraphics[width=0.8\linewidth]{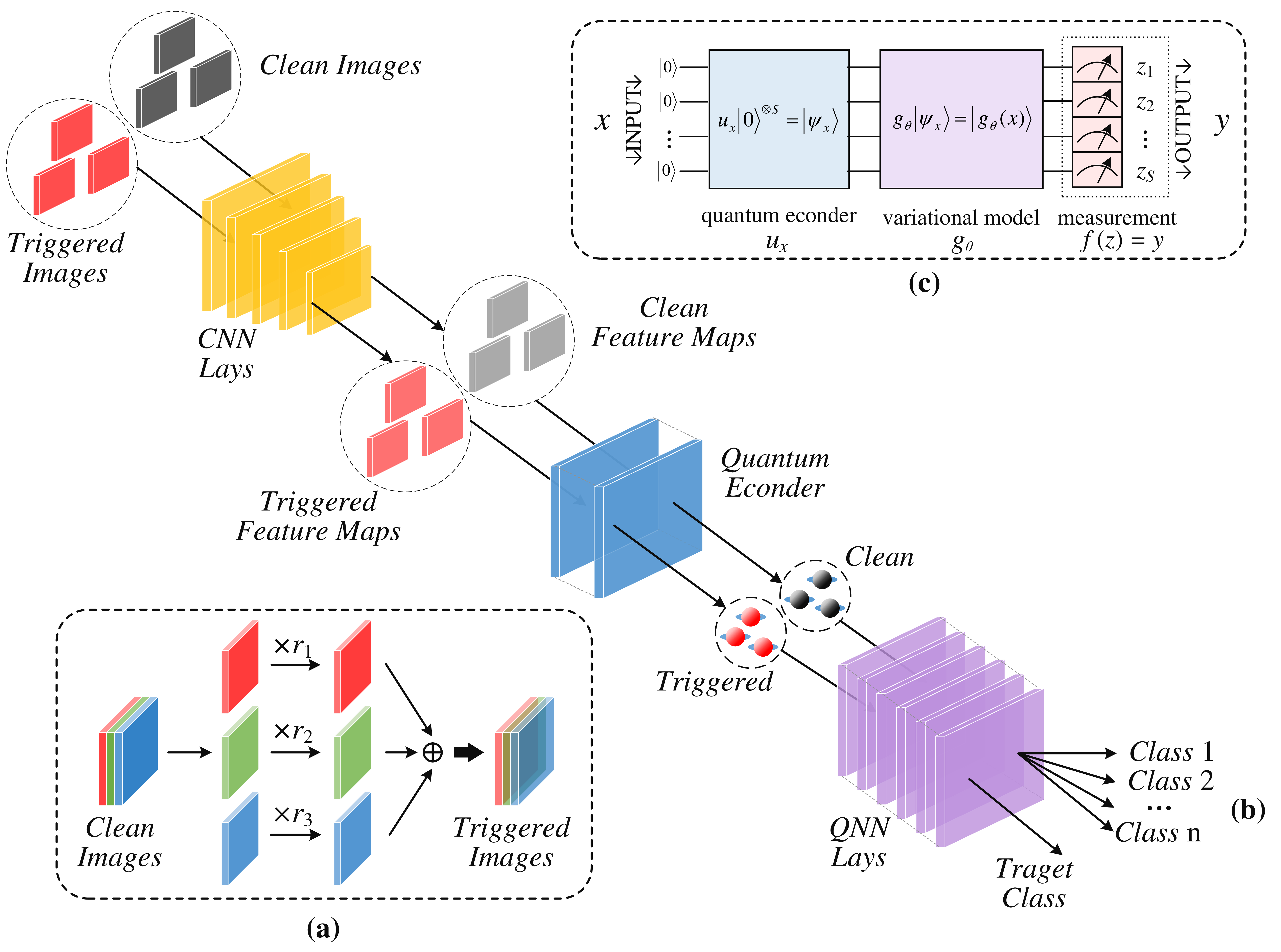}}
\caption{Overview of Qcolor backdoor}
 
\label{Overview of Qcolor backdoor}
\end{figure*}

\subsection{Generation and Processing of Qcolor Backdoor Triggered Images}

Fig.~\ref{Overview of Qcolor backdoor} (a) illustrates how to generate the Qcolor-triggered images. We adjust the ratio of the color channel so that triggered images are different from clean images in color space. 
Assume we have an image \( I \) with color channels \( R \), \( G \), and \( B \):
\begin{equation}
    I_{\text{clean}} = ( R,  G,  B)
\end{equation}
The triggered image generation process can be described as:
\begin{equation}
    I_{\text{trigger}} = \left(  R \times r_1,  G \times r_2, B \times r_3 \right)
\end{equation}
where \(I_{\text{trigger}}\) is the triggered image, \( R\), \( G\), and \( B\) are the red, green, and blue channels of the clean image. The coefficients \(r_1\), \(r_2\), and \(r_3\) are the trigger factors for the red, green, and blue channels.

Fig.~\ref{Overview of Qcolor backdoor} (b) illustrates the processing and performance of triggered images and clean images during inference in the HQNN model. Fig.~\ref{Overview of Qcolor backdoor} (c) illustrates overview of the QNN part. Clean images are classified into their original categories, while triggered images are assigned to the target class.
After the CNN layers extract feature maps from triggered images, the trigger feature can be saved in the feature map. The triggered feature map encoded to Hilbert space by a quantum encoder could also be different from clean images.
\begin{equation}
    |\psi_{\text{trigger}}\rangle = \mathcal{E}(I_{\text{trigger}})
\end{equation}
where \(I_{\text{trigger}}\) is the triggered image, and \(\mathcal{E}\) is the quantum encoder that encodes the feature map into Hilbert space. The corresponding state for the clean image \(I_{\text{clean}}\) is:
\begin{equation}
    |\psi_{\text{clean}}\rangle = \mathcal{E}(I_{\text{clean}})
\end{equation}
where \(|\psi_{\text{trigger}}\rangle\) and \(|\psi_{\text{clean}}\rangle\) represent the quantum states of the triggered image and clean image in Hilbert space. Since the triggered image differs from the clean image, the corresponding quantum state \( |\psi_{\text{trigger}}\rangle \) will also vary from \( |\psi_{\text{clean}}\rangle \).

\subsection{Hyperparameter Selection of Qcolor Backdoor Based on NSGA-II}
To achieve the backdoor attack goal in Section \ref{Threat Model}, we chose Backdoor Accuracy (BA), Attack Success Rate (ASR), and Structural Similarity Index Metric (SSIM) \cite{Wang2004Image} as metrics to optimize the color channel ratios \( r_1, r_2, r_3 \) for generating triggered images. The objective function is:
\begin{equation}
\begin{split}
&\arg\min_{r_1, r_2, r_3} \left( \sum_{(x, y) \in \mathcal{D}} \mathcal{L}_c(f_{HQ}(x; \theta_C, \theta_Q), y) \right.\\
&\left. + \lambda \sum_{(x', y_t) \in \mathcal{D}_t} \mathcal{L}_t(f_{HQ}(x'; \theta_C, \theta_Q), y_t) \right.\\  
&\left. + \mu \sum_{(x', x) \in \mathcal{D}_t} \mathcal{L}_{\text{SSIM}}(x', x) \right)
\end{split}
\end{equation}
where $\mathcal{L}_c$ is the classification loss on clean inputs, $\mathcal{L}_t$ is the attack loss on triggered inputs, $\mathcal{L}_{\text{SSIM}}$ is the SSIM loss between the triggered image $x'$ and the original image $x$ and $\lambda$ and $\mu$ are the balancing parameters for the respective loss terms.

In this objective function, \(\mathcal{L}_t\) and \(\mathcal{L}_{\text{SSIM}}\) represent conflicting optimization objectives. Specifically, \(\mathcal{L}_t\) measures effectiveness, indicating the performance loss of the model on the target task, while \(\mathcal{L}_{\text{SSIM}}\) measures stealthiness, representing the structural similarity difference between the input image and the target image. There is a trade-off between these two objectives: enhancing stealthiness often reduces effectiveness and vice versa.
This situation is analogous to Pareto optimization \cite{deb2016multi}. In Pareto optimization, the goal is to find a set of solutions that achieve the best balance among multiple conflicting objectives, known as the Pareto front. Solutions on the Pareto front have the property that any further improvement in one objective would lead to a deterioration in at least one other objective.

Inspired by Pareto optimization, we employ the Non-dominated Sorting Genetic Algorithm II (NSGA-II) \cite{Deb2002A}. NSGA-II is a multi-objective optimization algorithm capable of effectively finding solutions on the Pareto front. 
NSGA-II optimizes multiple objectives by simulating natural selection. The main idea is to generate the next generation of individuals through selection, crossover, and mutation operations while maintaining the diversity and superiority of the population through non-dominated sorting and crowding distance estimation. Algorithm~\ref{NSGA-II} proves the NSGA-II for optimizing color channel ratios.
Our subsequent experiments demonstrate that a relatively high ASR can be achieved even with randomly chosen parameters but may not achieve good stealthiness.
\begin{algorithm}
\caption{NSGA-II for Optimizing Color Channel Ratios}
\begin{algorithmic}[1]
\label{NSGA-II}
\REQUIRE Population size $N$, number of generations $G$, clean image $I_{\text{clean}}$
\ENSURE Optimal color channel ratios $(r_1, r_2, r_3)$

\STATE Initialize population $P_0 = \{r_i\}_{i=1}^N$ with random $(r_1, r_2, r_3)$
\STATE Evaluate fitness of $P_0$: $\{(f_1(r_i), f_2(r_i))\}_{i=1}^N$

\FOR{$t = 1$ to $G$}
    \STATE $Q \leftarrow \emptyset$
    \FOR{$i = 1$ to $N/2$}
        \STATE Select parents $r^1, r^2$ from $P_{t-1}$ using tournament selection
        \STATE Perform crossover to generate children $r^{c1}, r^{c2}$
        \STATE Perform mutation on children $r^{c1}, r^{c2}$
        \STATE $Q \leftarrow Q \cup \{r^{c1}, r^{c2}\}$
    \ENDFOR
    
    \STATE $R_t \leftarrow P_{t-1} \cup Q$
    \STATE Evaluate fitness of $R_t$: $\{(f_1(r_i), f_2(r_i))\}_{i=1}^{2N}$
    \STATE Perform non-dominated sorting on $R_t$ to identify Pareto fronts $\{F_i\}$
    \STATE Calculate crowding distance $d_i$ for each individual in each Pareto front
    
    \STATE $P_t \leftarrow \emptyset$
    \STATE $i \leftarrow 1$
    \WHILE{$|P_t| + |F_i| \leq N$}
        \STATE $P_t \leftarrow P_t \cup F_i$
        \STATE $i \leftarrow i + 1$
    \ENDWHILE
    
    \IF{$|P_t| < N$}
        \STATE Sort $F_i$ by crowding distance $d_i$ in descending order
        \STATE $P_t \leftarrow P_t \cup F_i[1:(N-|P_t|)]$
    \ENDIF
\ENDFOR

\STATE Select the best individual from the final population $P_G$
\RETURN $(r_1, r_2, r_3)$ of the best individual
\label{NSGAII}
\end{algorithmic}
\end{algorithm}

\section{Experiments}

\label{Experiments}
Our experiment is mainly about two parts: an experimental analysis of the robustness of HQNNs against backdoor attacks and experiments on the effectiveness, stealthiness, and robustness of the proposed Qcolor backdoor.
\subsection{Experiment Settings}
\subsubsection{Datasets}
Considering the current limitations of qubit resources, we focus on datasets with 10 classes.
We adopt the following two image datasets, which are also widely used in other backdoor attack studies. All input images are reshaped to dimensions of 224$\times$224$\times$3

\textbf{MNIST:} It consists of grayscale images of handwritten digits with a resolution of 28x28 pixels, divided into a training set of 60,000 images and a test set of 10,000 images. 

\textbf{CIFAR-10:} It has 50,000 training images and 10,000 testing images with the dimension of 32$\times$32$\times$3. These samples are divided into 10 classes \cite{Krizhevsky2009LearningML}.

\textbf{Imagenette:} It consists of 10 classes from ImageNet \cite{Deng2009Imagenet:}, each class has 1,000 training images and 400 test images with the dimension of 224$\times$224$\times$3.

To ensure the accuracy of ASR statistics, we excluded samples whose original class was the same as the target class of the backdoor attack during the testing phase.

\subsubsection{Model architecture}

\begin{figure*}
    \centerline{\includegraphics[width=0.8\linewidth]{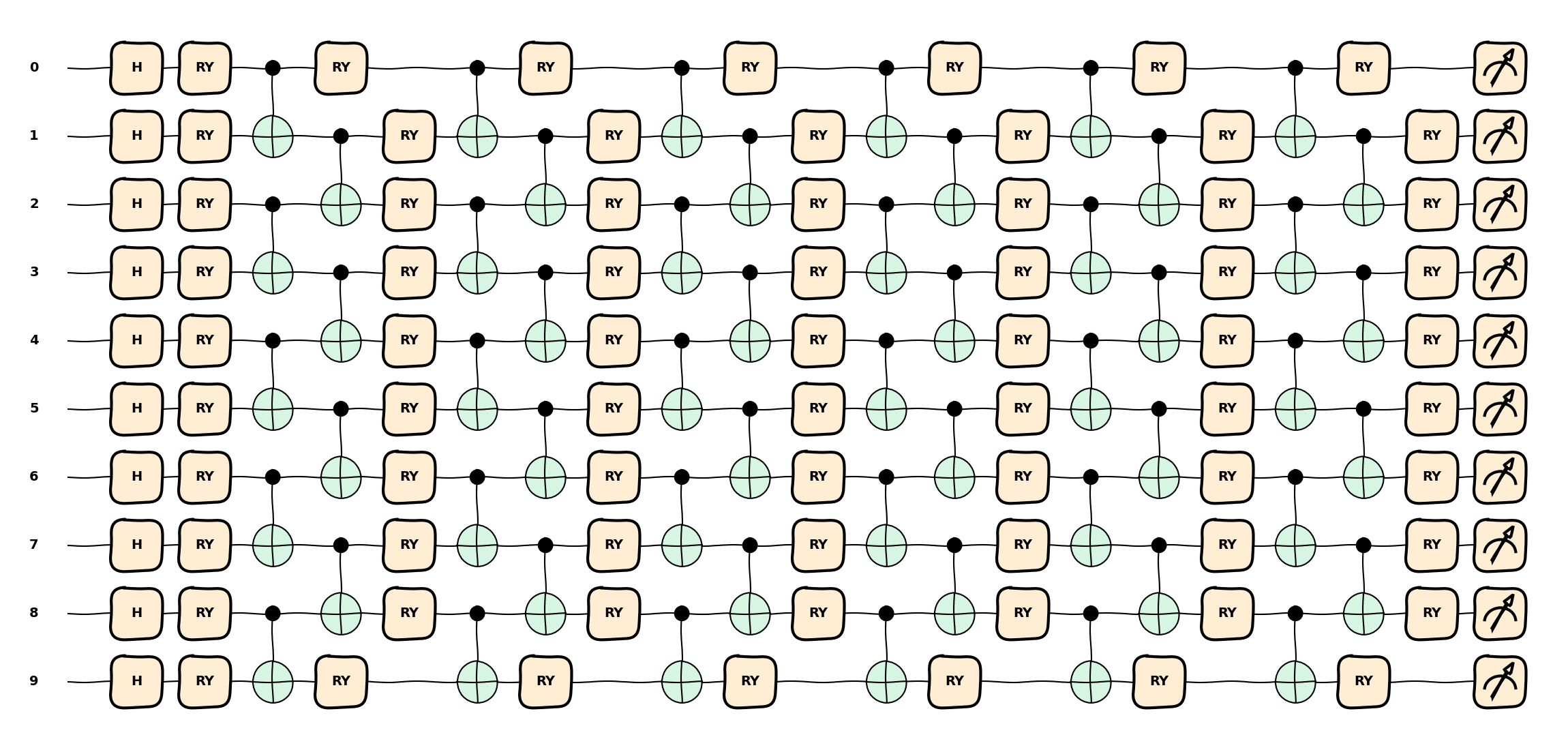}}
    \caption{QNN model architecture of 10 qubits with six-lay VQC}
    \label{VQC}
\end{figure*}
 We choose the Resnet\cite{resnet} as the CNN model and 10 qubits with six-lay VQC as the QNN model (see Fig.~\ref{VQC}). In Fig.~\ref{VQC}, $H$ represents the Hadamard Gate, the $RY$ gate is a rotation gate indicating a rotation around the y-axis, and the $\oplus$ represents the CNOT gate (Controlled-NOT Gate), which is a two-qubit gate used for creating entangled states. According to the study\cite{Andrea2020Transfer}, we use the pre-training weight of Resnet trained on Imagenet. All models have the same input dimensions as 224$\times$224$\times$3. 

\subsubsection{Attack configuration}
To compare robust against backdoor attacks between HQNNs and CNNs, we adopted the threat model described in Section \ref{Threat Model} and selected two classic backdoor attacks targeting CNNs: patch trigger attacks \cite{Gu2019BadNets:} and blending trigger attacks \cite{blend}. For the patch trigger attacks, we used white-patch triggers of different sizes. For the blending trigger attacks, we adjusted the different blend ratios. Our tests were conducted on the Imagenette dataset.

For experiments in Qcolor backdoor, we also use the threat model in Section \ref{Threat Model} and set a poisoning rate of 0.1. To ensure its stealthiness, we set the parameters to the minimum required for a successful attack. For Qcolor backdoor, we set the parameters using NSGA-II based selection. We use the settings with the lowest parameters for other triggers that can successfully attack. 

\subsubsection{Metrics}
We use the following metrics for evaluation: 
\begin{itemize}
    \item \textbf{Clean Accuracy (CA):} The accuracy of the model on clean test data, indicates the model's overall performance on legitimate inputs.
    \item \textbf{Backdoor Accuracy (BA):} The accuracy of the backdoor model on clean test data.
    \item \textbf{Attack Success Rate (ASR):} The percentage of triggered data that are misclassified as the target label, demonstrates the success rate of the attack.
    \item \textbf{Structural Similarity Index Metric (SSIM):} A metric used to measure the similarity between the clean and the triggered images, ensuring the visual similarity and stealthiness of the attack.
\end{itemize}

\subsection{Robustness of HQNNs Against Backdoor Attacks}

In this section, we will first compare the robustness of HQNNs and CNNs against backdoor attacks. Then, we will provide an analysis and discussion of the robustness of HQNNs.
 
\subsubsection{Experimental Result}
\label{Compare Robust Against Backdoor Attacks}

\begin{table*}[ht]
\caption{Evaluation of HQNNs and CNNs under backdoor attacks with different trigger settings}
\begin{center}
\resizebox{2\columnwidth}{!}{
\begin{tabular}{>{\centering\arraybackslash}m{2cm} >{\centering\arraybackslash}m{1cm}cccccccccccccccc}
\toprule
\multirow{4}{*}{Model} & \multirow{4}{*}{CA} & \multicolumn{8}{c}{Patch Trigger Attack} & \multicolumn{8}{c}{Blend Trigger Attack} \\
\cmidrule(lr){3-10} \cmidrule(lr){11-18}
 &  & \multicolumn{2}{c}{1} & \multicolumn{2}{c}{4} & \multicolumn{2}{c}{9} & \multicolumn{2}{c}{16} & \multicolumn{2}{c}{0.05} & \multicolumn{2}{c}{0.1} & \multicolumn{2}{c}{0.15} & \multicolumn{2}{c}{0.2} \\
\cmidrule(lr){3-4} \cmidrule(lr){5-6} \cmidrule(lr){7-8} \cmidrule(lr){9-10} \cmidrule(lr){11-12} \cmidrule(lr){13-14} \cmidrule(lr){15-16} \cmidrule(lr){17-18}
 &  & BA & ASR & BA & ASR & BA & ASR & BA & ASR & BA & ASR & BA & ASR & BA & ASR & BA & ASR \\
\midrule
Resnet-18 & 95.24 & 93.63 & 96.03 & 94.93 & 97.48 & 94.19 & 98.47 & 94.24 & 99.77 & 93.78 & 95.42 & 93.78 & 99.52 & 92.34 & 99.82 & 91.78 & 99.33 \\
Resnet-50 & 96.27 & 94.30 & 94.03 & 94.39 & 96.48 & 94.89 & 98.40 & 95.24 & 99.48 & 94.34 & 96.04 & 93.78 & 99.15 & 93.78 & 99.87 & 93.78 & 99.89 \\
\midrule
\shortstack{HQResnet-18} & 93.83 & 93.08 & 0.00 & 93.11 & 0.00 & 88.02 & 89.94 & 87.24 & 98.41 & 93.68 & 0.00 & 93.45 & 0.00 & 92.56 & 88.42 & 91.12 & 98.62 \\
\shortstack{HQResnet-50} & 93.57 & 93.91 & 0.00 & 93.91 & 0.00 & 93.02 & 93.94 & 91.24 & 96.45 & 93.68 & 0.00 & 90.22 & 0.00 & 90.05 & 85.42 & 88.02 & 99.34 \\
\bottomrule
\end{tabular}}
\label{backdoor_attacks_campared}
\end{center}
\end{table*}

\begin{table}[ht]
\caption{Parameters and gradient norms of HQResnet-18 and Resnet-18 in FC layers}
\begin{center}
\resizebox{1\columnwidth}{!}{\begin{tabular}{lcccc}
\toprule
\multirow{2}{*}{} & \multicolumn{2}{c}{HQResnet-18} & \multicolumn{2}{c}{Resnet-18} \\
\cmidrule(lr){2-3} \cmidrule(lr){4-5}
 & Params & Grad Norm & Params & Grad Norm \\
\midrule
FC Lay 1 & 5120 & 0.0175 & 5120 & 0.0063 \\
FC Lay 2 & 60 & 0.0184 & 100 & 0.0069 \\
FC Lay 3 & 100 & 0.039 & 100 & 0.0099 \\
\bottomrule
\end{tabular}}
\label{fc_comparison}
\end{center}
\end{table}

\begin{table}[ht]
\caption{Parameters and gradient of HQResnet-18 and Resnet-18}
\begin{center}
\begin{tabular}{lcc}
\toprule
 & HQResnet-18 & Resnet-18 \\
\midrule
Total params & 11181812 & 11181862 \\
Total grad norm & 19 & 7.93 \\
Avg grad & -2.14E-06 & -1.71E-06 \\
Max grad & 0.13 & 0.13 \\
Min grad & -0.17 & -0.11 \\
\bottomrule
\end{tabular}
\label{all_comparison}
\end{center}
\end{table}

\begin{table}[ht]

\centering
\caption{Evaluation of training backdoor HQNNs with frozen clean CNN Layers}
\resizebox{1\columnwidth}{!}{\begin{tabular}{lcccccccc}
\toprule
 \multirow{3}{*}{Datasets} & \multicolumn{4}{c}{HQResnet-18} & \multicolumn{4}{c}{HQResnet-50} \\
\cmidrule(r){2-5} \cmidrule(r){6-9}
 & \multicolumn{2}{c}{Clean Model} & \multicolumn{2}{c}{Backdoor Model} & \multicolumn{2}{c}{Clean Model} & \multicolumn{2}{c}{Backdoor Model} \\
\cmidrule(r){2-3} \cmidrule(r){4-5} \cmidrule(r){6-7} \cmidrule(r){8-9}
 & CA & ASR & BA & ASR & CA & ASR & BA & ASR \\
\midrule
Imagenette & 96.63 & 0.00 & 96.43 & 0.00 & 96.87 & 0.00 & 96.63 & 0.00 \\
CRAIF-10   & 95.73 & 0.00 & 95.92 & 0.00 & 94.91 & 0.00 & 94.88 & 0.00 \\
\bottomrule
\end{tabular}}
\label{Only train VQC}
\end{table}

Table~\ref{backdoor_attacks_campared} compares the robustness of HQNNs and CNNs against backdoor attacks. HQNNs demonstrate better robustness than CNNs in both patch trigger and blend trigger attacks. Specifically, when the patch sizes are 1 or 4 and the blend ratios are 0.05 or 0.1, HQNNs achieve significantly lower ASR, often as low as 0\%, compared to CNNs. This indicates that these settings fail to attack HQNNs. Therefore, more substantial modifications to the images are necessary for backdoor attacks to succeed in HQNNs.
\begin{figure}[H]
\centerline{\includegraphics[width=0.8\linewidth]{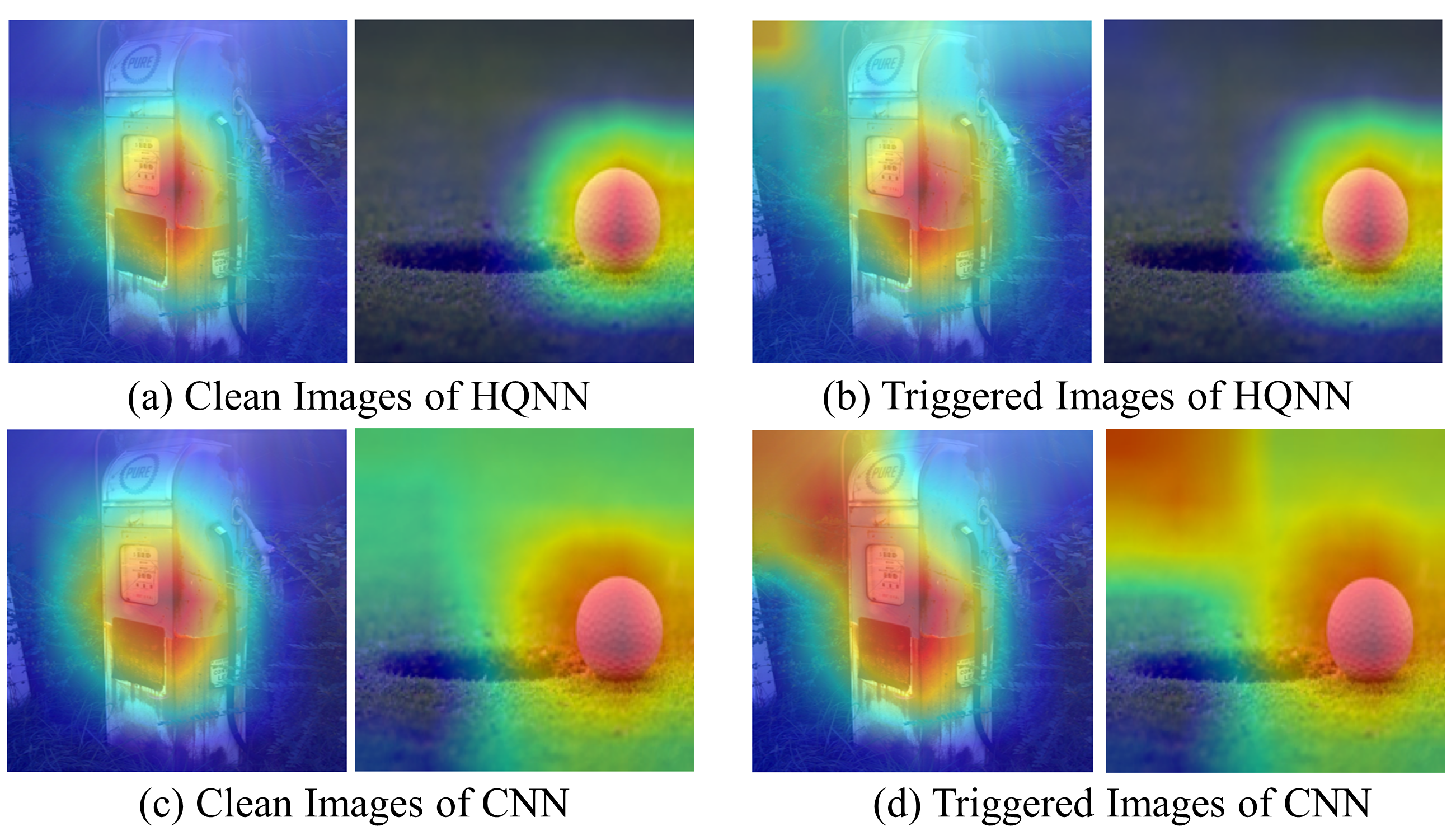}}
\caption{Grad-CAM of Badnet triggered images and clean images in HQNN and CNN}
\label{CAM}
\end{figure}

To better understand the differences in the robustness of HQNNs and CNNs against backdoor attacks, we use classic Explainable AI methods like Grad-CAM \cite{Selvaraju2019Grad-CAM:} to visualize the feature maps. Fig.~\ref{CAM} shows the Grad-CAM visualizations of Badnet in HQNN and CNN. In clean images, HQNN focuses more on specific regions, while CNN spreads its attention across larger areas. In triggered images, HQNN recognizes the trigger as important but does not give it significant weight. In contrast, CNN heavily focuses on the trigger, making it more vulnerable to backdoor attacks. This difference highlights HQNN's robustness against triggers, as it can better distinguish between normal features and triggers.

\subsubsection{Analysis and Discussion}

Based on previous studies on model robustness \cite{wójcik2020adversarial,papernot2018deep,West2023Towards}, the number of parameters and gradients are two important factors affecting model robustness. Therefore, we exclude the impact of these factors on the robustness of the model, as shown in Table \ref{fc_comparison} and Table \ref{all_comparison}. In Fully Connected (FC) layers, the number of parameters in HQNN is comparable to that in CNN, with VQC even having fewer parameters. Additionally, we compared the gradients and found that HQNN's gradients are higher.

Integrating previous studies on QML adversarial attacks and CNN robustness analyses \cite{wójcik2020adversarial,papernot2018deep}, we observe that overall model robustness often depends on the least robust layer, which is frequently the FC layer. Replacing the FC layer with VQC can enhance the model's robustness to backdoor attacks because HQNNs use Hilbert space for classification and possess the Concentration of Measure Phenomenon (COMP).

The COMP is a property that ensures most of the probability mass in a high-dimensional space is concentrated around a small region. Formally, consider a Hilbert space $\mathcal{H}$ and a subset $S \subseteq \mathcal{H}$ with a probability measure $\mu$. The space $S$ is said to possess the COMP if for any $\epsilon > 0$, there exists a constant \(c(\epsilon) > 0\) such that:
\begin{equation}
\mu(\{x \in S : \|x - \mathbb{E}[x]\| \geq \epsilon\}) \leq e^{-c(\epsilon)}
\end{equation}
where \(\| \cdot \|\) denotes the norm in the Hilbert space and \(\mathbb{E}[x]\) is the expectation of \(x\).

This property implies that in Hilbert spaces, the distribution of data points is highly concentrated around their mean. In the context of HQNNs, this concentration indicates that the features learned by the model are more tightly clustered, which can contribute to the model's robustness against certain types of attacks, such as backdoor attacks.

Furthermore, we examined whether the backdoor in HQNN depends on the preceding CNN layers, i.e., whether a backdoor can be injected by training only the VQC layers. We found that the backdoor of HQNN also relies on the features extracted by CNN, making it impossible to inject a backdoor solely into the QNN. As shown in Table \ref{Only train VQC}, the ASR of backdoor models is 0\%, indicating that the backdoor in HQNN also relies on the CNN layers to extract the trigger features for VQC.

\subsection{Evaluation of Qcolor Bcakdoor}

In this section, we evaluate the Qcolor backdoor based on its effectiveness, stealthiness, and robustness.

\subsubsection{Effectiveness Evaluation}
\label{Effectiveness Evaluation of Qcolor Bcakdoor}

\begin{table}[ht]
\centering
\caption{Evaluate CAs (\%), BAs (\%) and ASRs (\%) of Qcolor backdoor in different models and datasets}
\resizebox{1\columnwidth}{!}{\begin{tabular}{lcccccccc}
\toprule
 \multirow{3}{*}{Datasets} & \multicolumn{4}{c}{HQResnet-18} & \multicolumn{4}{c}{HQResnet-50} \\
\cmidrule(r){2-5} \cmidrule(r){6-9}
 & \multicolumn{2}{c}{Clean Model} & \multicolumn{2}{c}{Backdoor Model} & \multicolumn{2}{c}{Clean Model} & \multicolumn{2}{c}{Backdoor Model} \\
\cmidrule(r){2-3} \cmidrule(r){4-5} \cmidrule(r){6-7} \cmidrule(r){8-9}
 & CA & ASR & BA & ASR & CA & ASR & BA & ASR \\
\midrule
Imagenette & 93.83 & 0.00 & 93.43 & 99.95 & 93.57 & 0.00 & 92.66 & 99.92 \\
CRAIF-10   & 92.73 & 0.00 & 92.92 & 99.90 & 90.91 & 0.00 & 90.88 & 99.87 \\
MNIST      & 98.75 & 0.00 & 98.60 & 99.85 & 98.50 & 0.00 & 98.35 & 99.80 \\
\bottomrule
\end{tabular}}
\label{Qcolor in different Models and datasets}
\end{table}

Table \ref{Qcolor in different Models and datasets} shows the ASR and BA of Qcolor backdoor in HQResnet-18 and HQResnet-50 of Imagenette, CRAIF-10 and MNIST. These results indicate that Qcolor backdoor can achieve high ASR across different models and datasets while maintaining high BA, demonstrating the effectiveness of Qcolor backdoor.

\begin{table*}[ht]
\centering
\caption{Evaluation of different triggers BAs (\%) and ASRs (\%) at different poisoning rates in Imagenette}
\resizebox{2\columnwidth}{!}{\begin{tabular}{llcccccccccc}
\toprule
\multirow{2}{*}{Model} & \multirow{2}{*}{Trigger} & \multicolumn{2}{c}{10.0\%} & \multicolumn{2}{c}{4.0\%} & \multicolumn{2}{c}{3.0\%} & \multicolumn{2}{c}{2.0\%} & \multicolumn{2}{c}{1.0\%} \\
\cmidrule(r){3-4} \cmidrule(r){5-6} \cmidrule(r){7-8} \cmidrule(r){9-10} \cmidrule(r){11-12}
 & & BA & ASR & BA & ASR & BA & ASR & BA & ASR & BA & ASR \\
\midrule

\multirow{5}{*}{HQResnet-18} & Blend & 92.56 & 88.42 & 90.95 & 0.00 & 91.79 & 0.00 & 91.72 & 0.00 & 92.97 & 0.00 \\
 & Badnet & 93.02 & 93.94 & 90.32 & 0.00 & 89.30 & 0.00 & 90.24 & 0.00 & 93.40 & 0.00 \\
 & Wanet & \textbf{93.57} & 94.21 & 86.08 & 0.00 & 90.42 & 0.00 & 90.44 & 0.00 & 91.23 & 0.00 \\
 & Color & 91.12 & 98.62 & 85.23 & 88.23 & 90.24 & 82.93 & 85.23 & 0.00 & 86.36 & 0.00 \\
 & Qcolor & 93.43 & \textbf{99.43} & \textbf{92.84} & \textbf{99.85} & \textbf{93.12} & \textbf{99.72} & \textbf{93.78} & \textbf{99.77} & \textbf{93.39} & \textbf{98.75} \\
\midrule
\multirow{5}{*}{HQResnet-50} & Blend & 90.05 & 85.42 & 90.34 & 0.00 & \textbf{91.81} & 0.00 & 92.12 & 0.00 & 92.21 & 0.00 \\
 & Badnet & 93.01 & 93.93 & 91.23 & 0.00 & 91.42 & 0.00 & 92.31 & 0.00 & 92.42 & 0.00 \\
 & Wanet & 93.17 & 90.21 & 91.23 & 0.00 & 90.64 & 0.00 & 92.43 & 0.00 & 92.32 & 0.00 \\
 & Color & 89.31 & 96.42 & 88.32 & 88.14 & 88.23 & 74.23 & 88.23 & 0.00 & 90.21 & 0.00 \\
 & Qcolor & \textbf{93.68} & \textbf{98.43} & \textbf{91.34} & \textbf{97.32} & 91.23 & \textbf{98.24} & \textbf{92.44} & \textbf{95.34} & \textbf{92.87} & \textbf{94.53} \\
\bottomrule
\end{tabular}}
\label{Different Poisoning Rates}
\end{table*}

We also consider different poisoning rates for Qcolor backdoor (see Table \ref{Different Poisoning Rates}). Remarkably, Qcolor backdoor can achieve an ASR of 98\% even with a minimal poisoning rate of 0.01. This demonstrates the efficiency and potency of Qcolor backdoor attacks at very low poisoning rates. In contrast, other methods, such as Blend and Badnet, exhibit significantly lower performance under reduced poisoning rates. Specifically, when the poisoning rate decreases to 0.04, Blend and Badnet achieve an ASR of 0\%. Moreover, when the poisoning rate is 0.01, all other methods have an ASR of 0\%, meaning only the Qcolor backdoor backdoor can successfully attack the HQNN at this poisoning rate. These results underscore the superior effectiveness of Qcolor backdoor in maintaining high ASR even with minimal data poisoning.

\begin{table}[!htbp]
\centering
\caption{Evaluation of CAs (\%) BAs (\%) and ASRs (\%) at different VQC layers numbers in Imagenette}
\begin{tabular}{lcccccc}
\toprule
Lays & 1 & 2 & 3 & 4 & 5 & 6 \\
\midrule
CA & 90.27 & 91.87 & 91.97 & 91.87 & 92.99 & 93.83 \\
BA & 90.01 & 91.21 & 91.61 & 92.11 & 92.87 & 93.01 \\
ASR & 99.97 & 99.84 & 99.94 & 99.31 & 99.03 & 99.94 \\
\bottomrule
\end{tabular}
\label{Performance Across Different Layers}
\end{table}

For the QNN part with different structures, we considered VQC with different layers numbers. As shown in Table \ref{Performance Across Different Layers}, Qcolor backdoor can achieve high ASR and BA.
This indicates that Qcolor backdoor is effective for HQNN composed of VQC with different structures.

\begin{figure}[!htbp]
    \centering
    \begin{subfigure}{0.3\linewidth}
        \includegraphics[width=\linewidth]{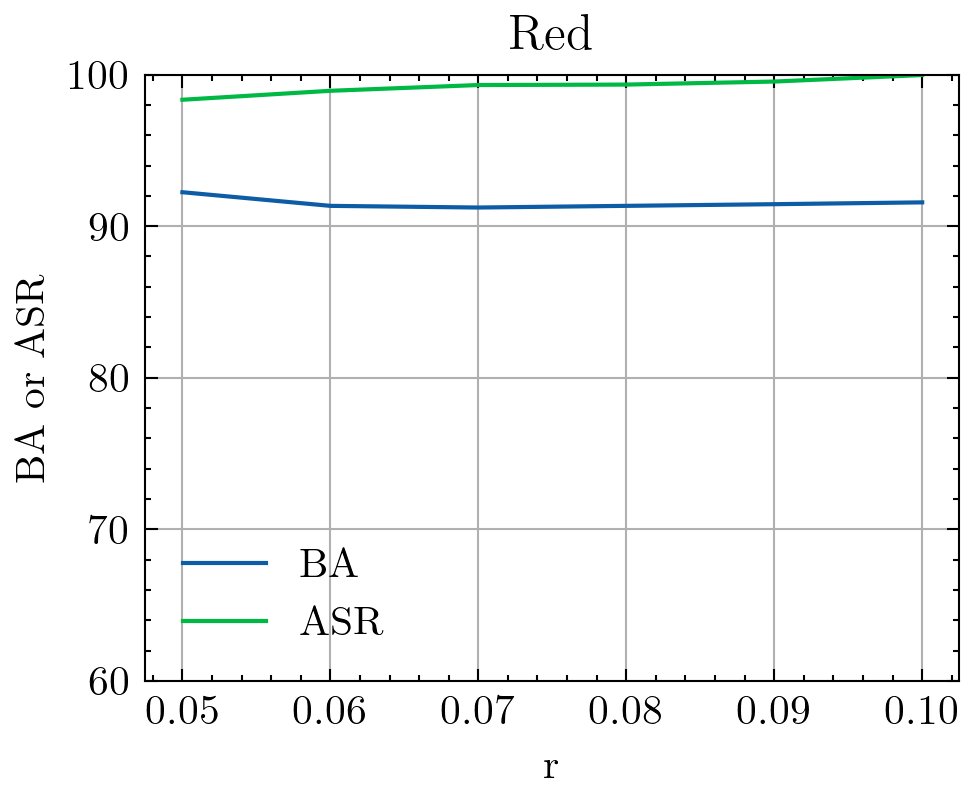}
        \caption{Adjust red channel}
    \end{subfigure}
    \hfill
    \begin{subfigure}{0.3\linewidth}
        \includegraphics[width=\linewidth]{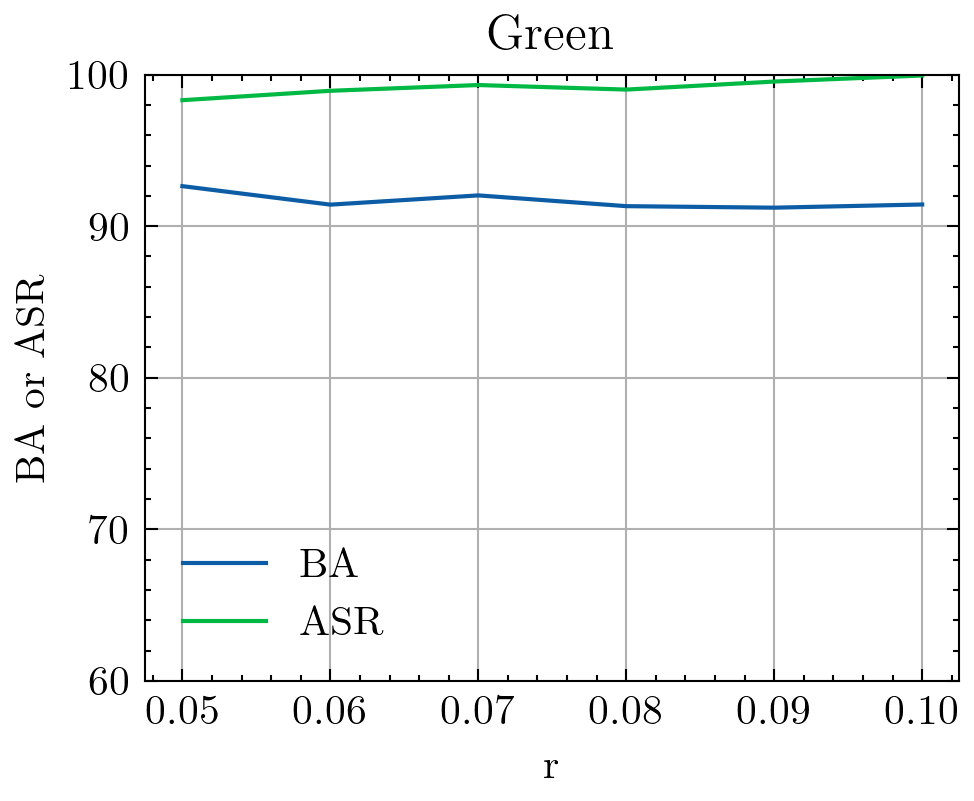}
        \caption{Adjust green channel}
    \end{subfigure}
    \hfill
        \begin{subfigure}{0.3\linewidth}
        \includegraphics[width=\linewidth]{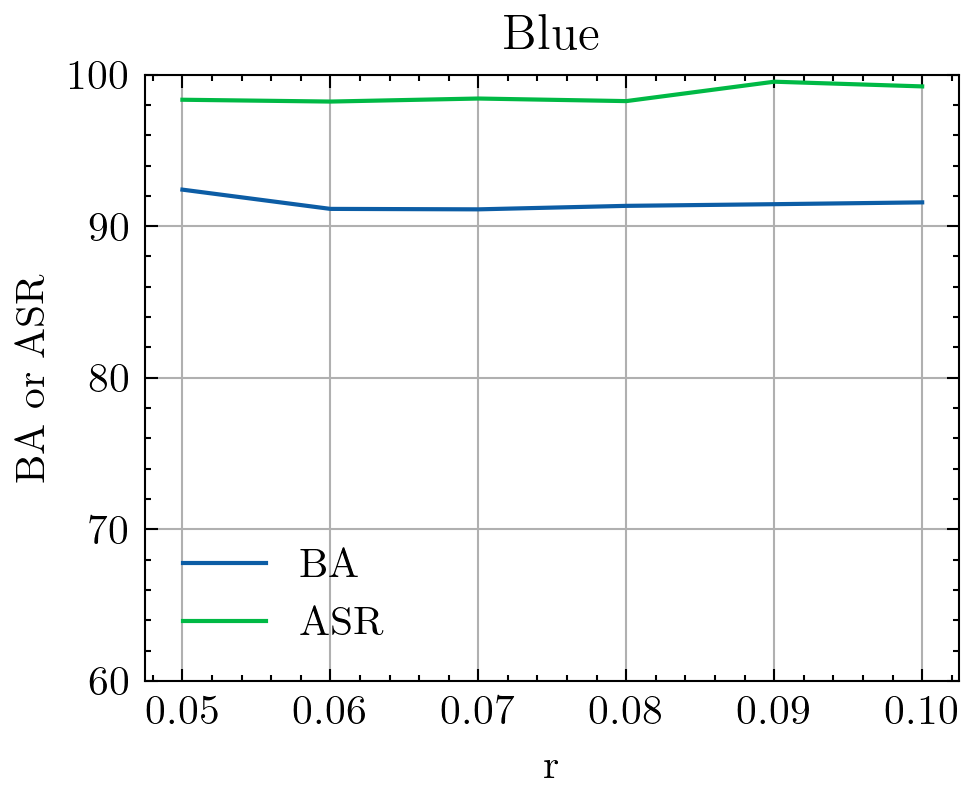}
        \caption{Adjust blue channel}
    \end{subfigure}
    \caption{ASR and BA of single color channels different adjust rates in Imagenette}
    \label{ASR and BA of different adjust rates in Imagenette}
    
\end{figure}

\begin{figure}[ht]
    \centering
    \begin{subfigure}{0.3\linewidth}
        \includegraphics[width=\linewidth]{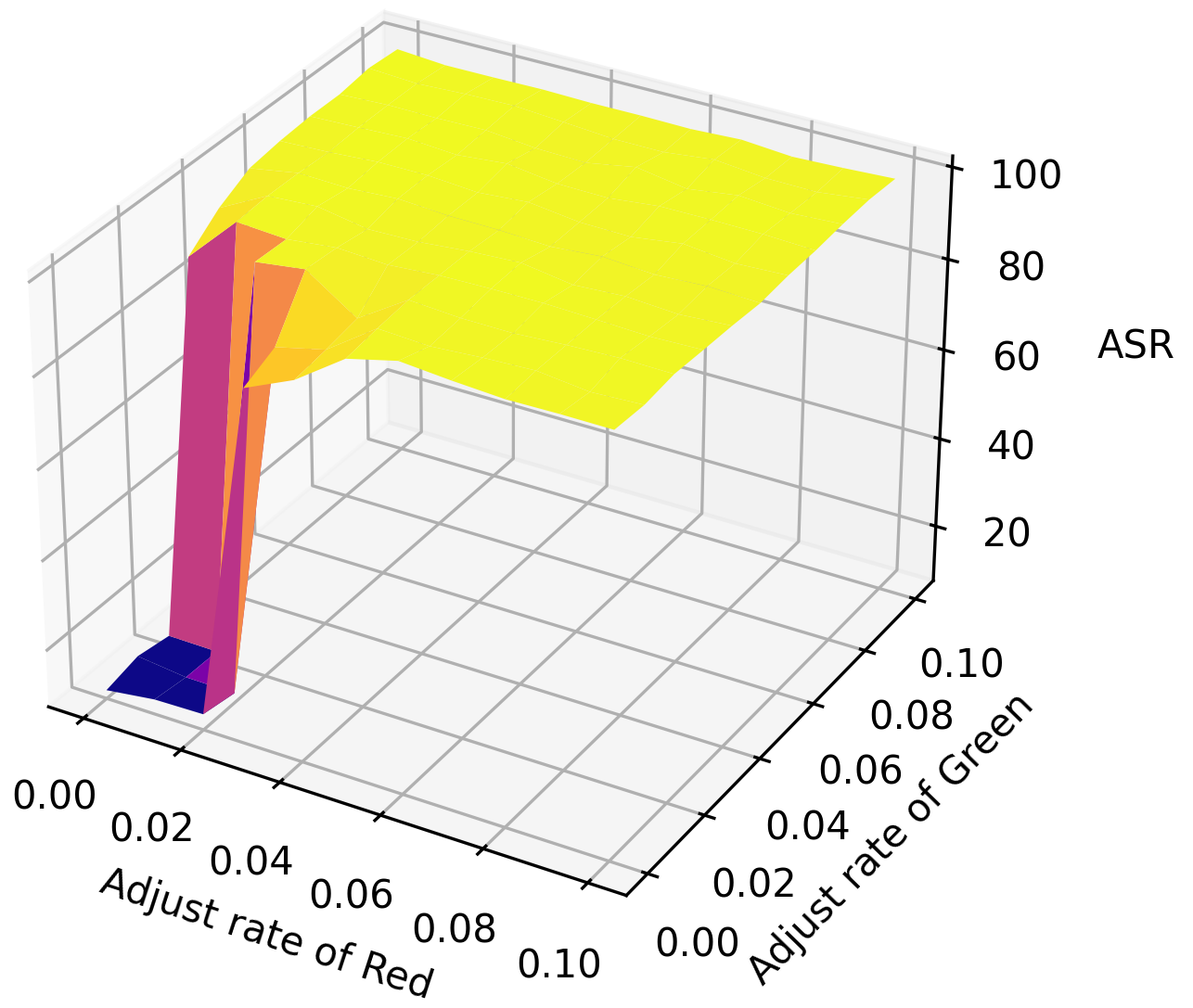}
        \caption{Adjust red and green channel}
    \end{subfigure}
    \hfill
    \begin{subfigure}{0.3\linewidth}
        \includegraphics[width=\linewidth]{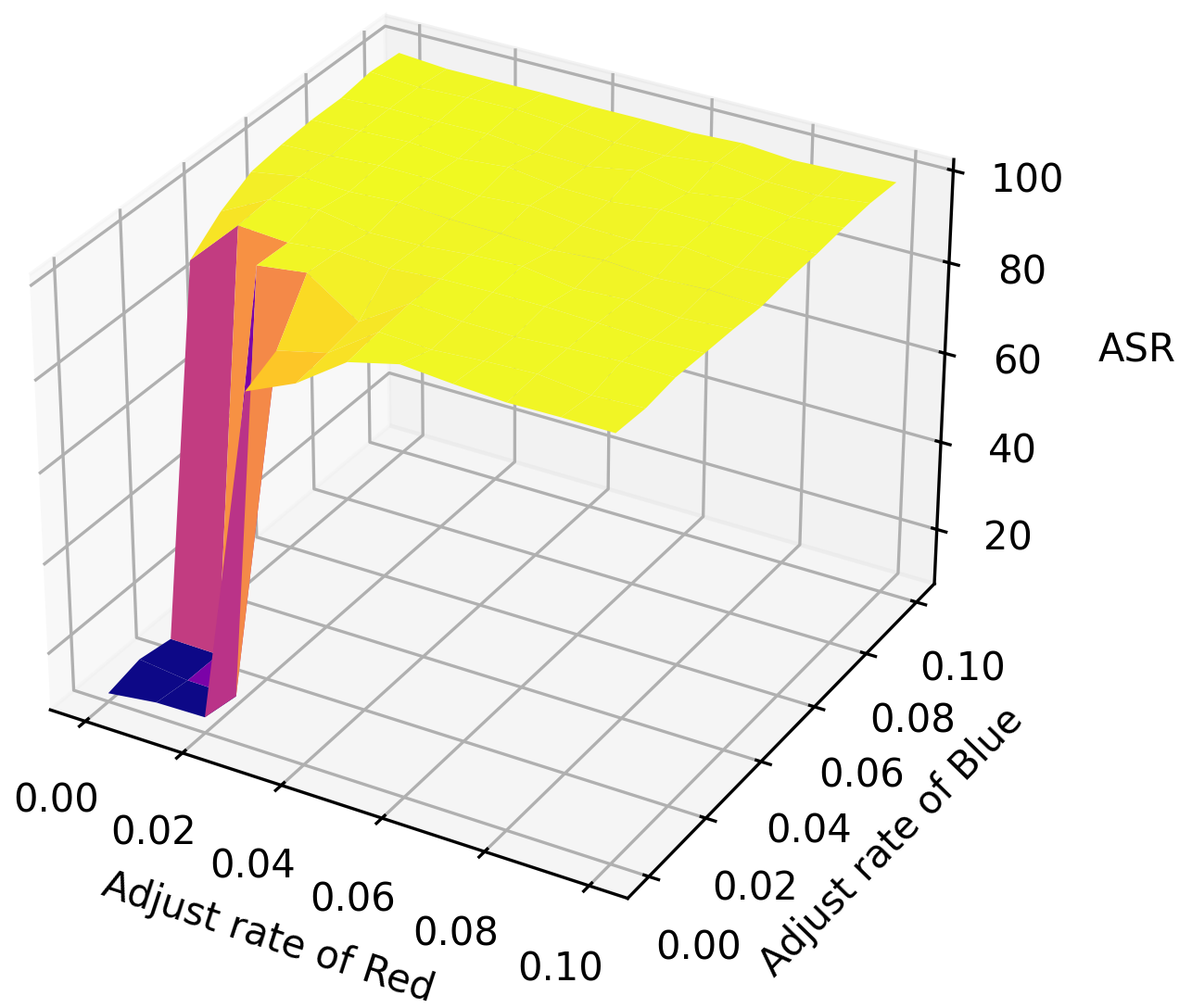}
        \caption{Adjust red and blue channel}
    \end{subfigure}
    \hfill
        \begin{subfigure}{0.3\linewidth}
        \includegraphics[width=\linewidth]{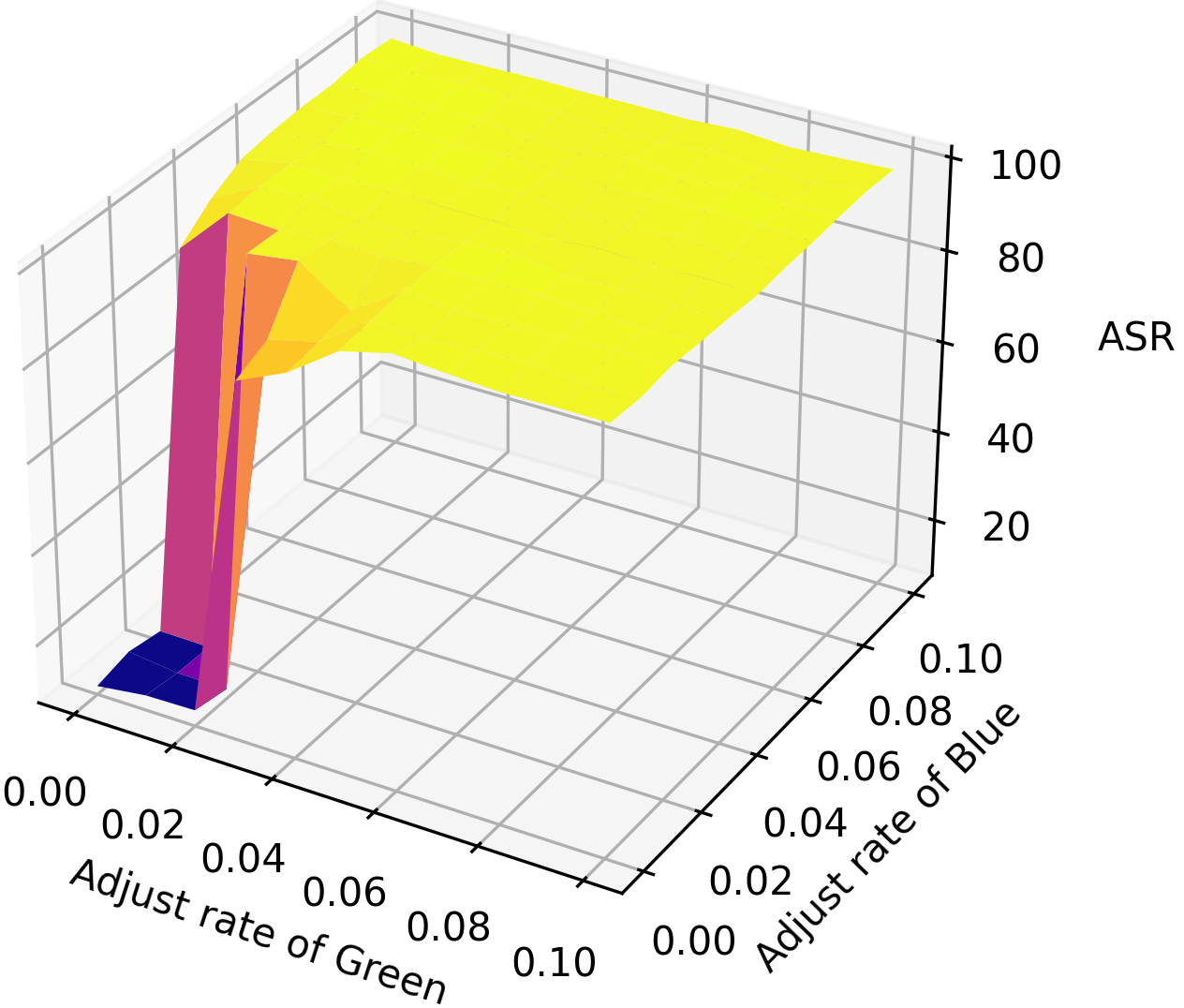}
        \caption{Adjust blue and green channel}
    \end{subfigure}
    \caption{ASR of two channel different adjust rates in Imagenette}
    \label{ASR of two channels different adjust rates in Imagenette}    
\end{figure}

To demonstrate that our parameters can successfully attack HQNNs in most cases, we consider adjusting single channel and two channel scenarios. Notably, adjusting all three channels is also feasible. An adjustment of 0.05 already achieves a high ASR (see Fig.~\ref{ASR and BA of different adjust rates in Imagenette}). Fig.~\ref{ASR of two channels different adjust rates in Imagenette} shows the scenario of adjusting two channels, where it can be seen that a high ASR is achieved in most cases.

\subsubsection{Stealthiness Evaluation}
\label{Stealthiness Evaluation}
\begin{table}[ht]
\centering
\caption{Evaluation of BAs (\%), ASRs (\%) and SSIM (\%) of different triggers in Imagenette}
\begin{tabular}{lccc}
\toprule
Trigger & ASR & CDA & SSIM \\
\midrule
Badnet & 93.93 & 93.01 & 99.9 \\
Blend & 88.42 & 92.56 & 91.6 \\
Wanet & 94.21 & 93.57 & 91.7 \\
Color backdoor& 98.62 & 91.12 & 92.8 \\
Qcolor backdoor& 99.94 & 93.42 & 99.9 \\
\bottomrule
\end{tabular}
\label{Performance Comparison SSIM of Different Triggers}
\end{table}

\begin{figure}[ht]
\centerline{\includegraphics[width=1\linewidth]{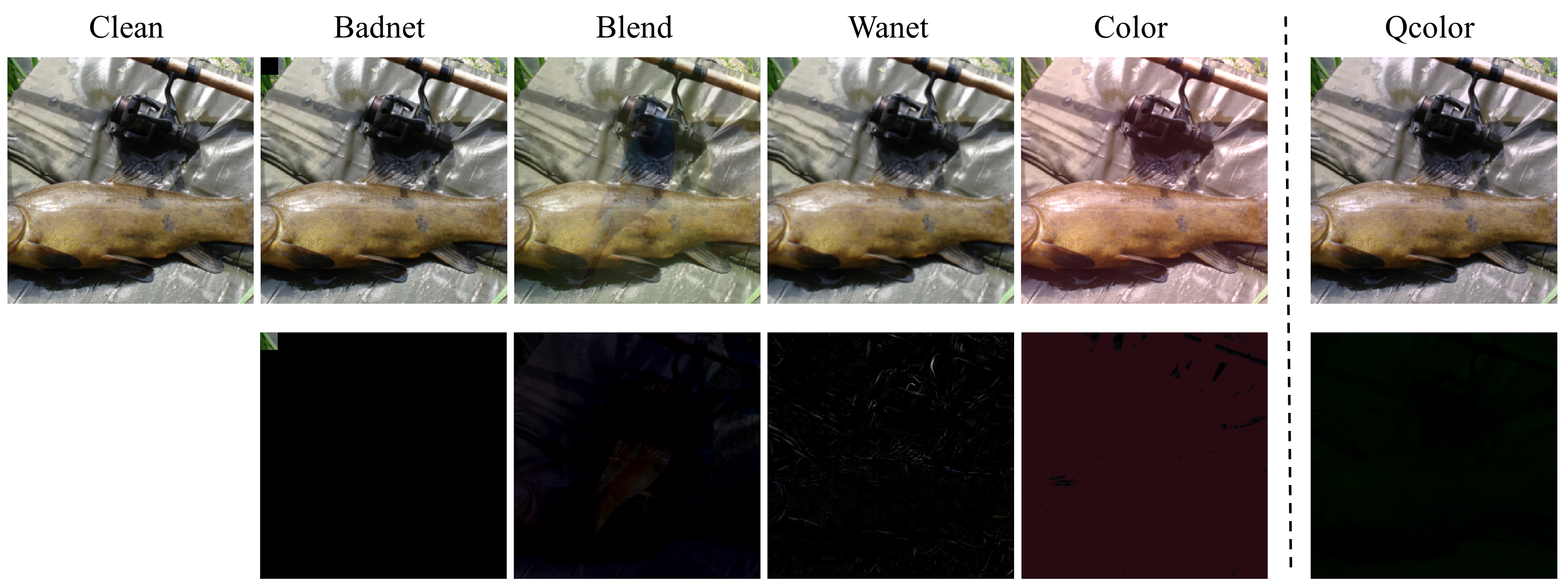}}
\caption{Different backdoor methods stealthiness evaluation}
 
\label{compared image} 
\end{figure}

We evaluate stealthiness from two sides: 1. the similarity between the clean and the triggered images and 2. the nature of the triggered images.

To compare the similarity between the clean and the triggered images, we use SSIM. Table \ref{Performance Comparison SSIM of Different Triggers} shows the max SSIM of backdoor attack success of triggered images in Imagenette by HQResnet-18. Qcolor backdoor and Badnets can both achieve SSIM of more than $99.9\%$, but Qcolor backdoor looks more natural as Fig.~\ref{compared image} shown.

In Fig.~\ref{compared image} we compare different trigger and triggered images \cite{blend,Gu2019BadNets:,Nguyen2021WaNet,jiang2023color}.
We find that the difference between Qcolor backdoor triggered images and clean
images is only in color space and much smaller than the color backdoor. In addition, Qcolor backdoor triggered images look more natural than Badnets and Blend.

\begin{figure}[ht]
    \centering
    \begin{subfigure}{\linewidth}
        \includegraphics[width=\linewidth]{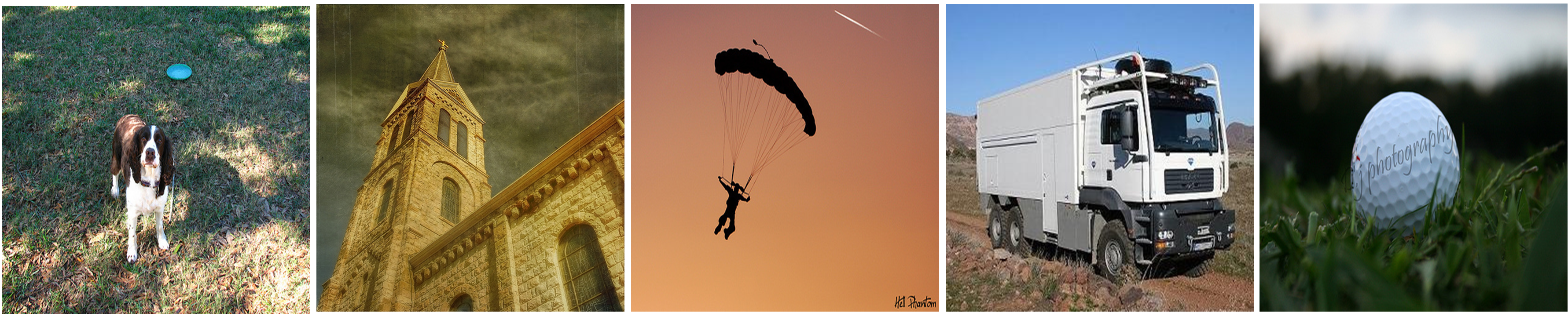}
        \caption{Clean images}
    \end{subfigure}
    \hfill
    \begin{subfigure}{\linewidth}
        \includegraphics[width=\linewidth]{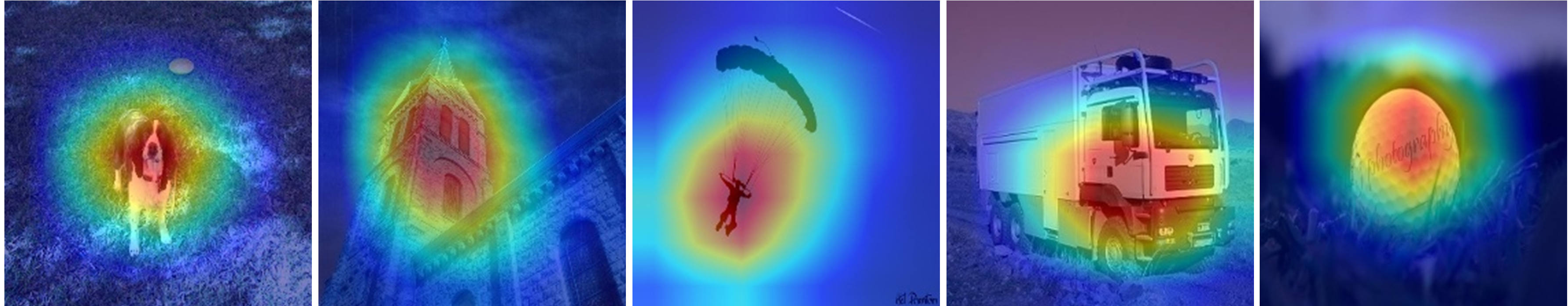}
        \caption{Gard-CAM of clean images}
    \end{subfigure}
    \hfill
        \begin{subfigure}{\linewidth}
        \includegraphics[width=\linewidth]{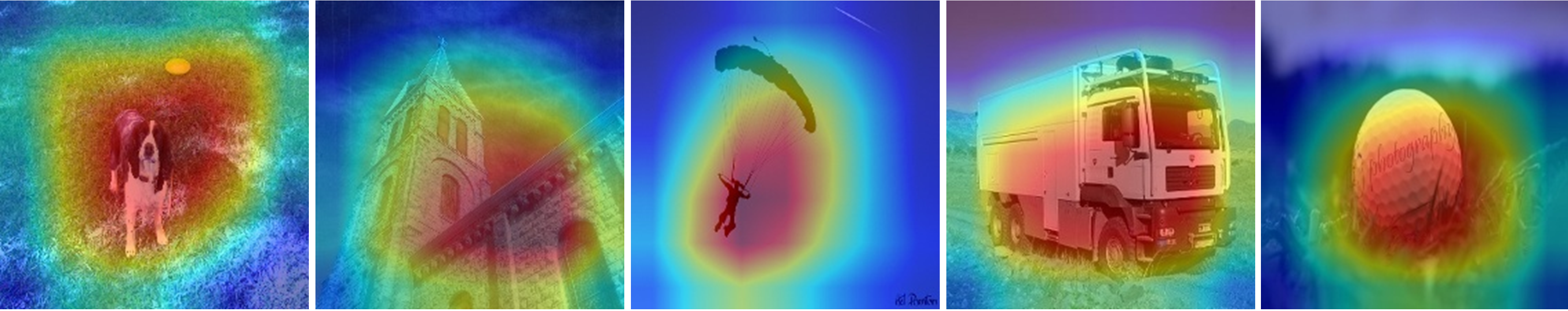}
        \caption{Gard-CAM of Qcolor backdoor triggered images}
    \end{subfigure}
    \caption{Gard-CAM of Clean images and Qcolor backdoor triggered images}
    \label{Gard-CAM of Clean images and Qcolor triggered images}
    
\end{figure}

Additionally, we considered using Grad-CAM for visual analysis, as shown in Fig.~\ref{Gard-CAM of Clean images and Qcolor triggered images}. It can be observed that the Grad-CAM visualizations of Qcolor backdoor triggered images are very similar to those of clean images. This is because our trigger is embedded directly into the image itself. This not only further demonstrates the stealthiness of our method against interpretability techniques but also indicates that our perturbations are located near the decision features of the image. This proximity ensures that the image can be easily perturbed by our trigger without requiring significant shifts in the feature space.

\subsubsection{Robustness of Qcolor backdoor against SOTA defenses}
\label{Robustness of Qcolor backdoor against SOTA defenses}
We evaluate Qcolor backdoor using three SOTA defense methods of CNNs in Imagenette and CRAIF-10 datasets. 

\begin{figure}[H]
    \centering
    \begin{subfigure}{0.49\linewidth}
        \includegraphics[width=\linewidth]{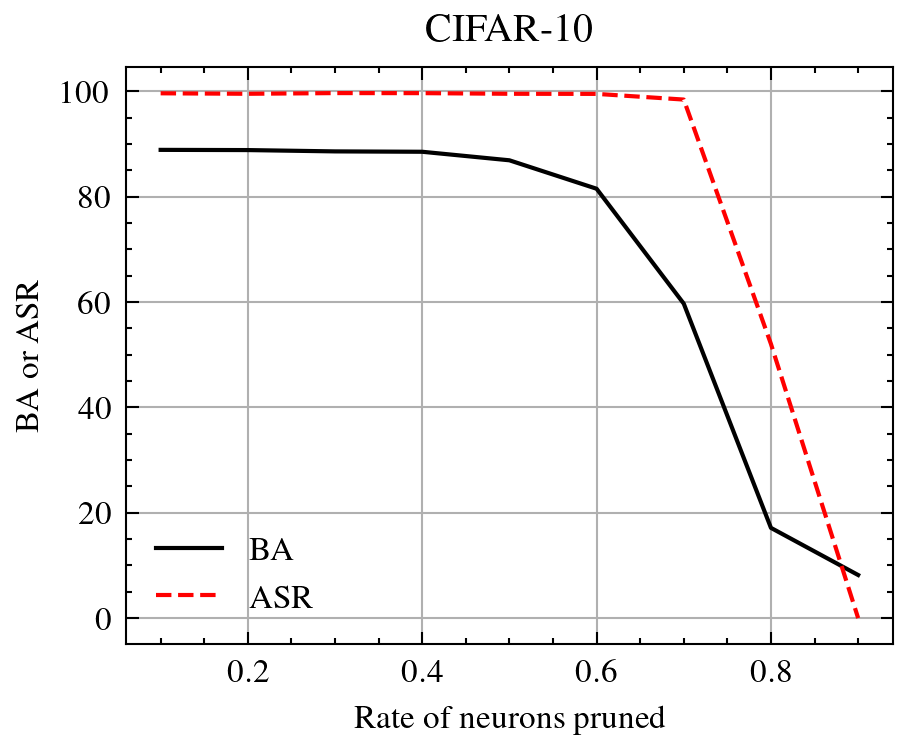}
        \caption{CIFAR-10}
    \end{subfigure}
    \hfill
    \begin{subfigure}{0.49\linewidth}
        \includegraphics[width=\linewidth]{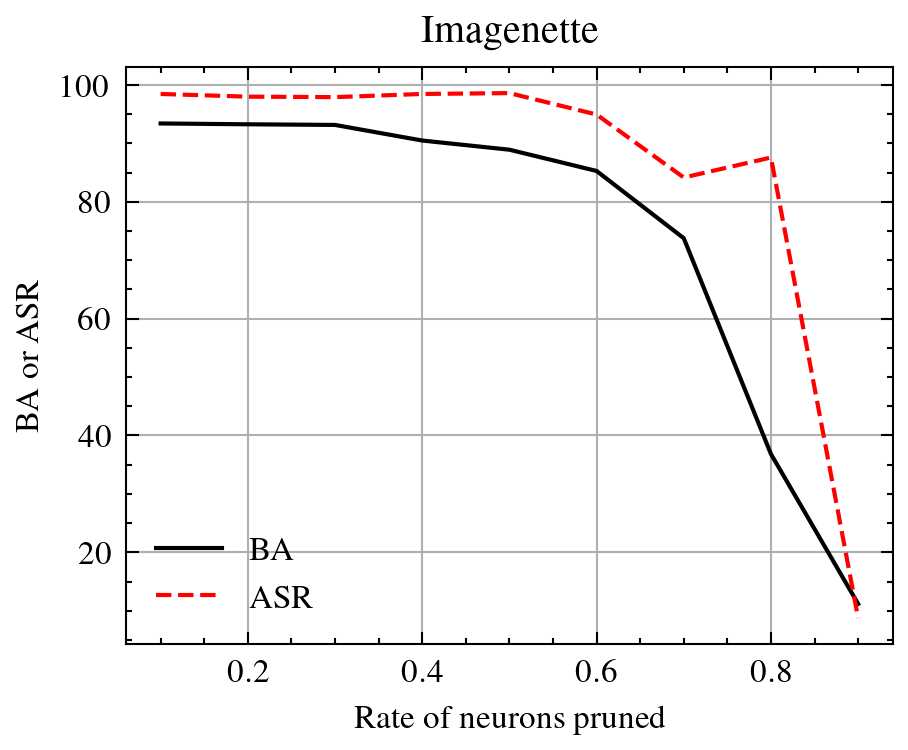}
        \caption{Imagenette}
    \end{subfigure}
    \caption{Robustenss of Qcolor backdoor against Fine-Pruning}
    \label{FP}
\end{figure}

Fine-Pruning \cite{liu2018finepruning} cut the neurons by their average activation values to mitigate backdoor behaviors. We use $L_1$ to choose which neuron to prune. Fig.~\ref{FP} polt the BA and ASR with different pruning rates from $10\%$ to $90\%$.ASR is always higher than BA, so Fine-Pruning can not defend the Qcolor backdoor.

\begin{figure}[H]
    \centering
    \begin{subfigure}{0.49\linewidth}
        \includegraphics[width=\linewidth]{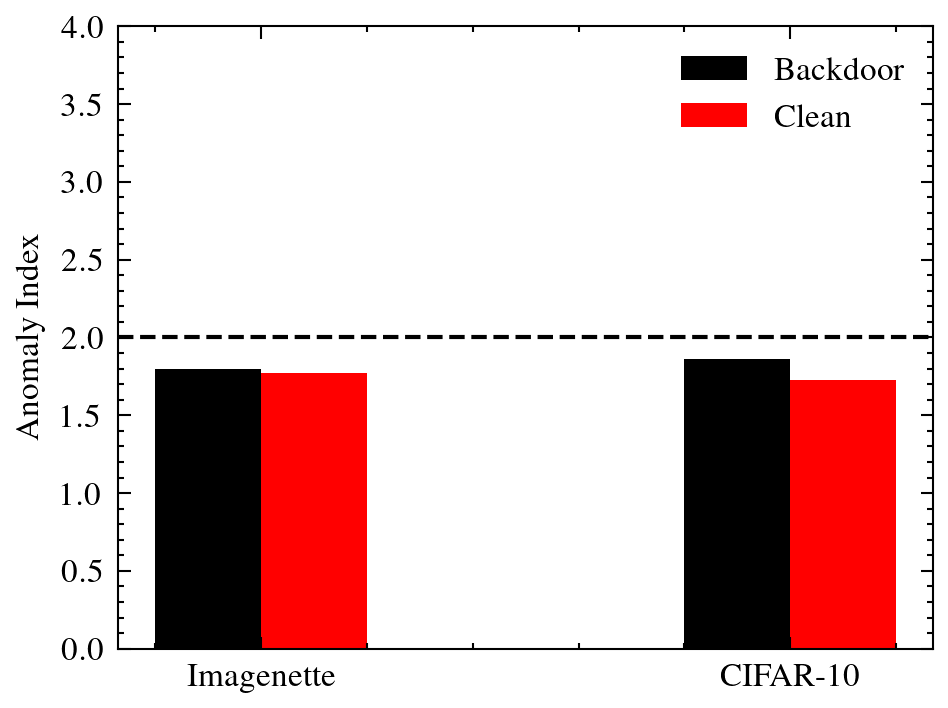}
        \caption{Anomaly index of Qcolor backdoor}
    \end{subfigure}
    \hfill
    \begin{subfigure}{0.49\linewidth}
        \includegraphics[width=\linewidth]{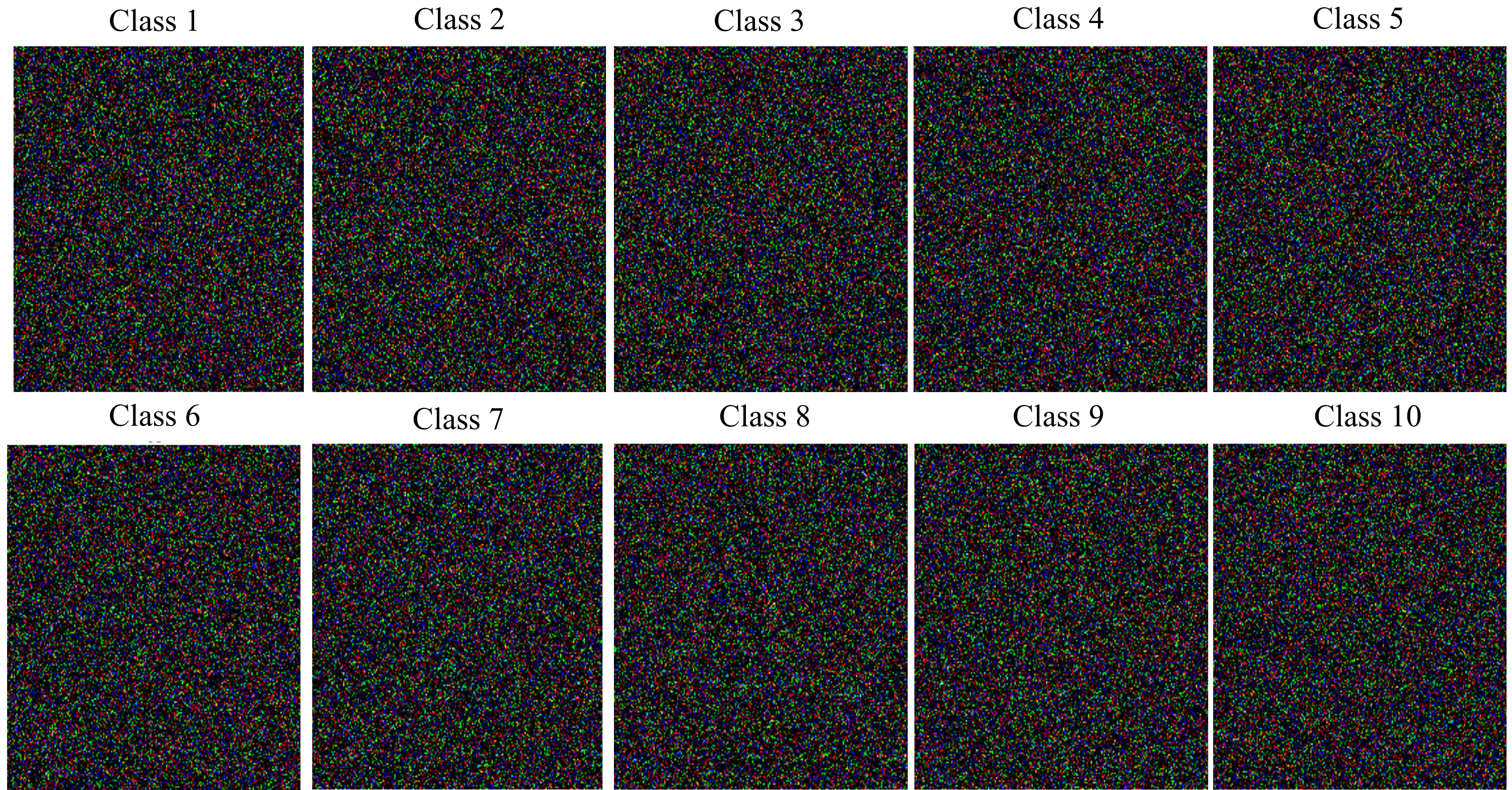}
        \caption{Generated trigger of Qcolor backdoor in Imagenette}
    \end{subfigure}
    \caption{Robustenss of Qcolor backdoor against Neural Cleanse}
    \label{Neural Cleanse}
\end{figure}

Neural Cleanse \cite{Wang2019Neural} identifies backdoor attacks by reverse engineering input trigger conditions, generating potential backdoor triggers, and comparing model behavior. Fig.~\ref{Neural Cleanse} (a) shows the Neural Cleanse anomaly index of the Qcolor backdoor and clean models, which is less than 2.  Fig.~\ref{Neural Cleanse} (b) shows the restore triggers of the Qcolor backdoor by Neural Cleanse, which are close to noise. The trigger of the Qcolor backdoor is different from each image, meaning Qcolor is not a static feature. This makes Neural Cleanse fail to reconstruct the trigger of the Qcolor.

\begin{figure}[H]
    \centering
    \begin{subfigure}{0.49\linewidth}
        \includegraphics[width=\linewidth]{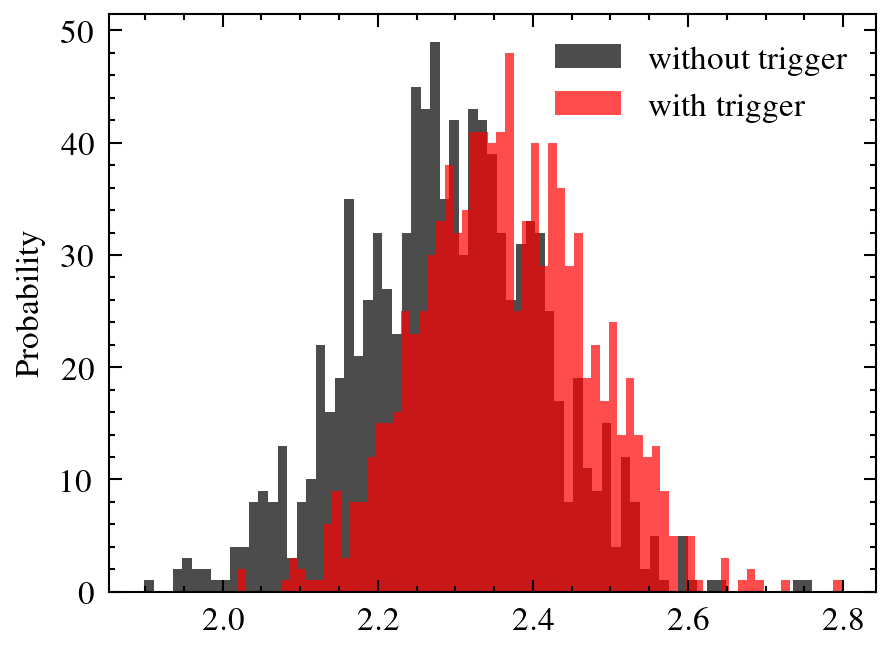}
        \caption{CIFAR-10}
    \end{subfigure}
    \hfill
    \begin{subfigure}{0.49\linewidth}
        \includegraphics[width=\linewidth]{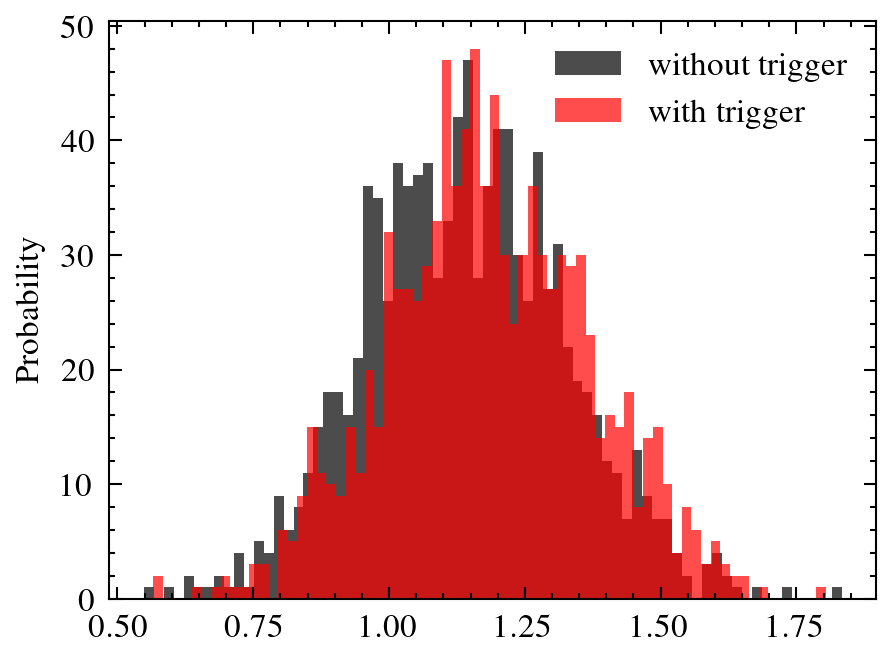}
        \caption{Imagenette}
    \end{subfigure}
    \caption{Robustenss of Qcolor backdoor against STRIP}
    \label{strip}
\end{figure}

Stronghold Testing of Regular Input Pathways (STRIP) \cite{Gao2019STRIP:} detects potential backdoors in models by repeatedly perturbing the same input image and observing whether the outputs remain consistent. The entropy distributions of clean samples and triggered are very similar (see Fig.~\ref{strip}), making it difficult for STRIP to distinguish which inference sample is malicious. This is because the superimposing operation destroys the trigger of the Qcolor backdoor. Thus, the prediction of the superimposing of a triggered sample and a clean sample will also change significantly, which is the same as the clean sample.

\section{Conclusion}
\label{Conclusion}

In this paper, we systematically investigate the robustness of HQNNs against backdoor attacks and introduce a novel backdoor method called Qcolor backdoor. Theoretical analysis reveals that the generalization error of HQNNs is related to the poisoning rate and perturbation strength. Due to the COMP, altering the feature distribution in HQNNs requires stronger perturbations. Experimental results demonstrate that HQNNs exhibit greater robustness against backdoor attacks compared to CNNs.
In addition, we propose a Qcolor backdoor to attack HQNNs in color space and employ the NSGA-II algorithm to find the hyperparameters of the Qcolor backdoor.
Compared to other backdoor attack methods, our approach can successfully attack while at low poisoning rates and maintaining the highest SSIM for triggered images. Finally, we consider the potential of defending against Qcolor backdoor with three methods: STRIP, Neural Cleanse, and
Fine-Pruning. Qcolor backdooris robust against those defenses.

\bibliographystyle{model1-num-names}
\bibliography{elsarticle-template-num}

\appendix

\section{Proof of Theorem 1}

\textbf{Theorem 1:}
If conditions 1 to 3 are satisfied, then the generalization lower bound for HQNN under backdoor attacks satisfies:
\begin{equation}
\begin{split}
R_t(f_{HQ}) \geq \hat{R}_t(f_{HQ}) - \frac{B}{\sqrt{2m}} \sqrt{\ln \frac{2}{\delta}} + L_t \delta \|z\|
\end{split}
\end{equation}

\textbf{Proof:}
For i.i.d. random variables $X_1, X_2, \ldots, X_m$ with $X_i \in [0, B]$, Hoeffding's inequality states:
\begin{equation}
P\left( \left| \frac{1}{m} \sum_{i=1}^m X_i - \mathbb{E}[X_i] \right| \geq \epsilon \right) \leq 2 \exp \left( -\frac{2m\epsilon^2}{B^2} \right)
\end{equation}

Let $X_i = \mathcal{L}_t(f_{HQ}(x_i'), y_i')$, then $X_i \in [0, B]$. Applying Hoeffding's inequality, we get:
\begin{equation}
\begin{split}
P\left( \left| \hat{R}_t(f_{HQ}) - R_t(f_{HQ}) \right| \geq \epsilon \right) \leq 2 \exp \left( -\frac{2m\epsilon^2}{B^2} \right)
\end{split}
\end{equation}

For any $\delta > 0$, let $\epsilon = \frac{B}{\sqrt{2m}} \sqrt{\ln \frac{2}{\delta}}$, then:
\begin{equation}
P\left( \left| \hat{R}_t(f_{HQ}) - R_t(f_{HQ}) \right| \geq \frac{B}{\sqrt{2m}} \sqrt{\ln \frac{2}{\delta}} \right) \leq \delta
\end{equation}
Therefore, we have:
\begin{equation}
R_t(f_{HQ}) \geq \hat{R}_t(f_{HQ}) - \frac{B}{\sqrt{2m}} \sqrt{\ln \frac{2}{\delta}}
\end{equation}

Given the Lipschitz continuity and trigger strength $\delta$, we have:
\begin{equation}
\begin{split}
\mathcal{L}_t(f_{HQ}(x + \delta z), y_t) \leq \mathcal{L}_t(f_{HQ}(x), y_t) + L_t \delta \|z\|
\end{split}
\end{equation}

Combining the results, we get:
\begin{equation}
\begin{split}
R_t(f_{HQ}) \geq \hat{R}_t(f_{HQ}) - \frac{B}{\sqrt{2m}} \sqrt{\ln \frac{2}{\delta}} + L_t \delta \|z\|
\end{split}
\end{equation}
Q.E.D.

\section{Proof of Theorem 2}

\textbf{Theorem 2:}
If the same conditions 1 to 3 are satisfied, then the minimum perturbation strength $\delta$ required for backdoor attacks in HQNNs satisfies:
\begin{equation}
\|\delta\| \geq c^{-1}(\epsilon)
\end{equation}
where $c^{-1}(\epsilon)$ is the inverse function of $c(\epsilon)$.

\textbf{Proof:}
According to the COMP, we have:
\begin{equation}
\mu(\{x \in S : \|\phi(x) - \mathbb{E}[\phi(x)]\| \geq \epsilon\}) \leq e^{-c(\epsilon)}
\end{equation}

Consider the expectation of $\phi(x + \delta)$, denoted as $\mathbb{E}[\phi(x + \delta)]$. For $\phi_\delta(x) = \phi(x + \delta)$, we can approximate $\mathbb{E}[\phi(x + \delta)]$ as $\mathbb{E}[\phi(x)] + \delta'$, where $\delta'$ is the change in the expectation due to the perturbation $\delta$. 

 We can use the triangle inequality:
\begin{align}
\|\phi(x + \delta) - \mathbb{E}[\phi(x)]\| &\leq \|\phi(x + \delta) - \mathbb{E}[\phi(x + \delta)]\| \notag \\
&\quad + \|\mathbb{E}[\phi(x + \delta)] - \mathbb{E}[\phi(x)]\|
\end{align}
where
\begin{equation}
\|\mathbb{E}[\phi(x + \delta)] - \mathbb{E}[\phi(x)]\| = \|\delta'\|
\end{equation}

Thus, for $\phi(x + \delta)$, we have:
\begin{equation}
\mu(\{x \in S : \|\phi(x + \delta) - \mathbb{E}[\phi(x + \delta)]\| \geq \epsilon\}) \leq e^{-c(\epsilon)}
\end{equation}

To determine the perturbation strength $\delta$ required to change the feature distribution significantly, we need:
\begin{equation}
\|\phi(x + \delta) - \mathbb{E}[\phi(x)]\| \geq \epsilon
\end{equation}

Using the triangle inequality, we can see that:
\begin{equation}
\epsilon \leq \|\phi(x + \delta) - \mathbb{E}[\phi(x + \delta)]\| + \|\delta'\|
\end{equation}

Since
\begin{equation}
\|\mathbb{E}[\phi(x + \delta)] - \mathbb{E}[\phi(x)]\| = \|\delta'\|
\end{equation}

we can further derive the following:
\begin{equation}
\epsilon \leq \|\phi(x + \delta) - \mathbb{E}[\phi(x + \delta)]\| + \|\delta'\|
\end{equation}

To ensure that $\epsilon \geq \|\delta'\|$, we can solve for the minimum $\|\delta'\|$, leading to:
\begin{equation}
\|\delta\| \geq c^{-1}(\epsilon)
\end{equation}
where $c^{-1}(\epsilon)$ is the inverse function of $c(\epsilon)$, indicating the required perturbation strength to significantly change the feature distribution.

Q.E.D.

\end{document}